\documentclass[reqno,11pt]{amsart}

\usepackage{fancyhdr}
\usepackage{datetime}
\settimeformat{ampmtime}
\usepackage[svgnames]{xcolor}
\usepackage{setspace}

\usepackage[svgnames]{xcolor}
\usepackage{pgf}
\usepackage{tikz}
\usepackage{mathpazo} 
\usepackage[scaled]{helvet} 
\usepackage{courier} 
\normalfont
\usepackage[T1]{fontenc}

\usetikzlibrary{patterns}
\usepackage{soul}
\usepackage{mathrsfs}
\usepackage{stmaryrd}

\usepackage[hidelinks=true]{hyperref}
\usepackage{hyperref}
\usepackage{natbib}
\usepackage{xfrac}
\usepackage[color=Orange]{todonotes}
\usepackage{layout}
\usepackage{microtype}
\usepackage{fancyhdr}
\usepackage{mathtools}

\usepackage{cleveref}

\newtheorem{example}{Example}
\newtheorem{proposition}{Proposition}
\newtheorem{corollary}{Corollary}
\newtheorem{lemma}{Lemma}
\newtheorem{definition}{Definition}

\newtheorem*{CRND}{Assumption R}
\newtheorem*{MEP}{Assumption M}

\newtheorem*{PN}{Assumption N}

\newtheorem*{MR-3}{Assumption S}

\newtheorem*{MI}{Assumption I}
\newtheorem*{SC}{Assumption SC}
\newtheorem*{SCP}{Assumption SCP}
\newtheorem*{ER}{Assumption ER}

\newtheorem*{MV}{Assumption V}

\newtheorem*{E}{Assumption E}

\newtheorem*{PR}{Assumption PRD}

\usepackage{amssymb}

\renewcommand{\Pr}{\mathbb{P}}

\DeclareMathOperator{\arginf}{arginf}
\DeclareMathOperator{\argsup}{argsup}

\newcommand\abstraction{
This paper studies the rationalization and identification of binary games where players have correlated private types. Allowing for correlation is crucial in global games and in models with social interactions as it represents  correlated information and homophily, respectively.  Our approach is fully nonparametric in the joint distribution of types and the strategic effects in the payoffs. First, under monotone pure Bayesian Nash Equilibrium strategy, we characterize all the restrictions if any on  the distribution of players' choices imposed by the game-theoretic model as well as restrictions associated with two assumptions frequently made in the empirical analysis of discrete games. Namely, we consider exogeneity of  payoff shifters relative to  private information, and mutual
independence of private information given payoff shifters.  Second, we study the nonparametric identification of the payoff functions 
and types distribution.
We show that  the model with exogenous payoff shifters is fully identified up to 
a single location--scale normalization  under some exclusion restrictions and rank conditions. 
Third,  we discuss partial identification under weaker conditions and multiple equilibria. Lastly, we briefly point out the implications of our results for model testing and estimation.\\ 

\noindent
\textbf{Keywords:} Rationalization, Identification, Discrete Game, Social Interactions, Global Games

\noindent
\textbf{JEL}: C14, C18, C31, C35, C52, and Z13
}

\usepackage{subfig}

\begin{document}

\thispagestyle{empty}

\title{\textsc{Rationalization and Identification of \\ Binary Games with Correlated Types}$^*$}

\thanks{$^*$We thank Victor Aguirregabiria, Andres Aradillas--Lopez, Emmanuel Guerre, Laurent Mathevet,  Bernard Salanie, Matthew Shum, Steven Stern, Elie Tamer, Xun Tang, Neil Wallace and Nese Yildiz for useful comments. We also thank seminar participants at Pennsylvania State University, Texas A\&M University, University of Chicago, Brown University, University of Texas at Austin,  Columbia University, Rice University, New York University, Shanghai University of Finance and Economics, and University of Toronto, as well as at Texas Metrics Camp 2013, the 2013 North American Summer Meeting of the Econometric Society at University of Southern California, and  the 2013 Asian Meeting of the Econometric Society at National University of Singapore. The second and third authors respectively acknowledge financial supports from the National Science Foundation through grant SES 1148149 and the Summer Research Fellowship of University of Texas.}

\author{
\href{mailto:nliu@psu.edu}{Nianqing Liu$^\star$}} 
\thanks{$^\star$School of Economics, Shanghai University of Finance and Economics, Shanghai, China, \href{mailto:nliu@shufe}{nliu@shufe.edu.cn}}
\author{
\href{mailto:qvuong@nyu.edu}{Quang Vuong$^\dag$}}
\thanks{$^\dag$(corresponding author) Department of Economics, New York University, 19 W. 4th Street, 6FL, New York, NY, 10012,
\href{mailto:qvuong@nyu.edu}{qvuong@nyu.edu}}
\author{
\href{mailto:h.xu@austin.utexas.edu}{Haiqing Xu$^{\ddag}$}}
\thanks{$^{\ddag}$Department of Economics, University of Texas at Austin, \href{mailto:h.xu@austin.utexas.edu}{h.xu@austin.utexas.edu}}

\date{\today}

\maketitle

\begin{abstract}
 \abstraction
\end{abstract}

\vspace{5ex}

\clearpage

\section{Introduction}
Many economic problems are naturally modeled as games of incomplete information \citep[see][]{morris2001global}.  Over the last decades, such games have been much successful for understanding the strategic interactions among agents in various economic and social situations. A leading example is auctions with e.g. \cite{vickrey1961counterspeculation}, \cite{riley1981optimal}, \cite{milgrom1982theory} for the theoretical side, and \cite{porter1995role}, \cite{guerre2000optimal} and \cite{athey2002identification} for the empirical component.  In this paper, we study the identification of static binary games of incomplete information where players have correlated types.\footnote{\cite{aradillas2014estimation} study identification and estimation of  ordered response games with independent types.  }   We characterize all the restrictions if any imposed by such games on the observables, which are the players' choice probabilities.   Following the work by \cite{laffont1996structural} and \cite{athey2007nonparametric} for auctions, our approach is fully nonparametric.

The empirical analysis of static discrete games is almost thirty years old.  
The range of applications includes labor force participation \cite[e.g.][]{bjorn1984simultaneous,bjorn1985econometric,kooreman1994estimation,soetevent2007discrete}, firms' entry decisions \citep[e.g.][]{bresnahan1990entry,BresnahanReiss1991a,berry1992estimation,tamer2003incomplete,berry2006identification,jia2008happens,ciliberto2009market}, and social interactions \citep[e.g.][]{kline2012identification}. These papers deal with discrete games under complete information. More recently,  discrete games under incomplete information have been used to analyze social interactions by \cite{brock2001discrete,brock2007identification} and \cite{xu2011social} among others, firm entry and location choices by \cite{seim2006empirical}, timing choices of radio stations commercials by \cite{sweeting2009strategic}, stock market analysts' recommendations by \cite{bajari2010estimating}, capital investment strategies by \cite{aradillas2010semiparametric}
and local grocery markets by \cite{grieco2011discrete}. This list is far from being exhaustive and does not mention the growing literature on estimating dynamic games. 

Our paper contributes to this literature in several aspects. First, we focus on  monotone pure strategy Bayesian Nash equilibria (BNE) throughout to bridge discrete game modeling with empirical analysis.  Monotonicity is a desirable property in many applications for both theoretical and empirical reasons. For instance, \cite{white2011causal} show that monotone strategies are never worse off than non--monotone strategies in a private value auction model. On theoretical grounds, \cite{athey2001single} provides seminal results on the existence of a monotone pure strategy BNE whenever a Bayesian game obeys the Spence--Mirlees single--crossing restriction. Relying on the powerful notion of contractibility,  \cite{reny2011existence} extends Athey's results and related results by \cite{mcadams2003isotone} to give weaker conditions ensuring the existence of a monotone pure strategy BNE. Using Reny's results, we establish the existence of a monotone pure strategy BNE under a weak monotonicity condition on the expected payoff in our setting. This condition is  satisfied  in most models used in the recent literature. For instance, in empirical IO, it is satisfied when the types are conditionally independent given payoff shifters. In social interaction games, it is also satisfied with  strategic complement payoffs and  positively regression dependent types.  Exceptions include \cite{aradillas2008identification, xu2010estimation}.
In our analysis, the importance of using
monotone pure strategy BNEs  lies in the fact that we can exploit (weak) monotonicity between observed actions and underlying types to identify nonparametrically the underlying game structure.  This  opens up the possibility of bringing some theoretical models such as global games \citep[see e.g.][]{carlsson1993global,morris1998unique} and models with social interactions \cite[see e.g. ][]{galeotti2010network} to nonparametric statistical inference.

Second, we allow players' private information/types to be correlated.   In finance and macroeconomics applications of global games \citep[e.g.  bank runs, currency crises, and bubbles; see][]{morris2001global},  private information are naturally positively correlated. See e.g. \cite{carlsson1993global,morris1998unique}.  In oligopoly entry, correlation among types allows us  to know ``whether entry occurs because of unobserved profitability that is independent of the competition effect'' \citep{berry2006identification}. In Sociology, correlation among players' types is crucial as it represents the ``Homophily'' phenomenon, which is the principle that people involved in interactions tend to be similar; see e.g. \cite{mcpherson2001birds,easley2010networks}.  The recognition of homophily in sociology has a long history: In the writings of Plato, for example, ``similarity begets friendship'' in his Phaedrus (360 BC). The homophily principle leads to friendship between people with similar demographics (age, race, education, etc) and with positively correlated types (taste, attitudes, etc). The former can be directly observed from the data and has been well documented in empirical sociology. Identifying the latter  is more challenging as it is unobserved to the researcher. It is worth pointing out that peer effects and homophily provide two complementary explanations for the common observation that friends tend to behave similarly.\footnote{In a linear social interaction model, \cite{manski1993identification} denotes them as endogenous effects and correlated effects, respectively.} Both of them can be separately identified  in our framework.  

In contrast, mutual independence of private information has been widely assumed in the empirical game literature.  See, e.g., \cite{brock2001discrete,pesendorfer2003identification,seim2006empirical,aguirregabiria2007sequential,sweeting2009strategic,bajari2010estimating,tang2010estimating,de2012inference,lewbel2012identification}. To our knowledge, the only exceptions are \cite{aradillas2010semiparametric}, \cite{wan2010semiparametric} and \cite{xu2010estimation}.  Such an independence of types is a convenient assumption, but imposes strong restrictions such as the mutual independence of players' choices given covariates, a property that is often invalidated by the data.\footnote{A model with unobserved heterogeneity and independent private information also generates dependence among players' choices conditional on covariates \citep[see e.g.][]{aguirregabiria2007sequential,grieco2011discrete}.  See also Section 5.4.}   
On the other hand, when private information is correlated, the BNE solution concept requires that each player's beliefs about rivals' choices depend on her private  information, thereby invalidating the usual two--step identification argument and estimation procedure,  see, e.g., \cite{bajari2010estimating}.  With such  type--dependent beliefs,   \cite{wan2010semiparametric} establish some upper/lower bounds for the beliefs in a semiparametric setting with linear--index payoffs.  Alternatively, \cite{aradillas2010semiparametric} adopts a different equilibrium concept related to \cite{aumann1987correlated}, in which each player's equilibrium beliefs do not rely on her private  information, but on her actual action. 

Third, our analysis is fully nonparametric in the sense that players' payoffs and the joint distribution of players' private information are subject to some mild smoothness conditions only.  As far as we know, with the exception of \cite{de2012inference} and \cite{lewbel2012identification}, every paper analyzing empirical discrete games has imposed parametric restrictions on the payoffs and/or the distribution of  private information. For instance, \cite{brock2001discrete,seim2006empirical,sweeting2009strategic} and \cite{xu2010estimation}  specify both payoffs and the private information distribution parametrically.  In a semiparametric context,   \cite{aradillas2010semiparametric,tang2010estimating} and  \cite{wan2010semiparametric} parameterize players' payoffs, while \cite{bajari2010estimating} parameterize the distribution of private information.  On the other hand, \cite{de2012inference} and \cite{lewbel2012identification} do not introduce any parameter but impose some restrictions on the payoffs' functional form.  In particular, they impose multiplicative separability in the strategic effect and assume that it is a known function (e.g. sum) of the other players' choices.     In a fully nonparametric setting,  our baseline discrete game model is the most general one and closest to that considered in game theory. We show that such a model imposes essentially no restrictions on the distribution of players' choices.  In other words, monotone pure strategy BNEs can explain almost all observed choice probabilities in discrete games.

In view of the preceding result, we consider the identification power and  model restrictions associated with two assumptions that are frequently made in the empirical analysis of discrete games. First, we consider the exogeneity of variables shifting  players' payoffs relative to players' private information, an assumption that has been frequently imposed in recent empirical work, e.g., \cite{brock2001discrete,seim2006empirical,sweeting2009strategic,aradillas2010semiparametric,bajari2010estimating,de2012inference} and \cite{lewbel2012identification}. We show that the resulting model restricts the distribution of players' choices conditional upon  payoff shifters and we characterize all those restrictions.  Specifically, the exogeneity assumption restricts the joint choice probability to be
a monotone function of the corresponding marginal choice probabilities.
Given the exogeneity assumption, we show that one can identify the equilibrium belief of the player at the margin under a mild support condition. We then characterize the partially identified set of payoffs and the distribution of private information under the exogeneity assumption and the support condition. The partially identified region is unbounded and quite large unless one imposes additional restrictions on the payoffs' functional form.

To achieve point identification, we consider some exclusion restrictions and rank conditions. We show that the copula function of the types' distribution is identified on an appropriate support. Then, the  players' payoffs  are identified up to scale for each fixed value of the exogenous state variables, as well as up to the marginal distributions of players' private information. Moreover, with a single location--scale normalization on the payoff function, we show that both the players' payoffs and distribution of types are fully identified. Our model can be viewed as an extension to a game theoretic setting of traditional  threshold--crossing models considered by, e.g., \cite{matzkin1992nonparametric}. An important difference is that the game setting allows us to exploit exclusion restrictions to achieve nonparametric identification of the distribution of errors. Such restrictions are frequently used in the empirical analysis of discrete games. See, e.g., \cite{aradillas2010semiparametric,bajari2010estimating,lewbel2012identification} and \cite{wan2010semiparametric}. 

For completeness, we consider a second assumption, namely the mutual independence of players' private information given payoff shifters. Specifically, we characterize all the restrictions imposed by exogeneity and mutual independence as considered almost exclusively in the empirical game literature.  We show that all the restrictions under this pair of assumptions can be summarized by the conditional mutual independence of players' choices given the payoff shifters.  In particular, we show that the restrictions imposed by mutual independence are stronger than those imposed by exogeneity  and monotonicity of the equilibrium.  In other words, any of the latter becomes redundant in terms of explaining players' choices  as soon as mutual independence and a single equilibrium are imposed.

The paper is organized as follows.  We introduce our baseline model in Section 2. We define and establish the existence of a monotone pure strategy BNE.     In Section 3, we study the restrictions imposed by the baseline model.  We also derive all the restrictions imposed by the exogeneity and mutual independence assumptions.  In Section 4, we  establish the nonparametric identification of the model primitives under some exclusion restrictions and rank conditions. In Section 5, we study the partial identification of the payoffs without exclusion restrictions.  We also discuss  three related issues: nonparametric estimation,  multiple equilibria in the DGP, and unobserved heterogeneity. Section 6 concludes with a brief discussion on testing  the model restrictions.

\section{Model and Monotone Pure Strategy BNE}

We consider a discrete game of incomplete information. There is a finite number of players, indexed by $i=1,2,\cdots,I$.  Each player simultaneously chooses a binary action $Y_i\in\{0,1\}$. Let $Y=(Y_1,\cdots, Y_{I})$ be an action profile and $\mathcal{A}=\{0,1\}^{I}$ be the space of  action profiles. Following standard convention, let $Y_{-i}$ and $\mathcal{A}_{-i}$ denote an action profile of all players except $i$ and the corresponding action profile space, respectively. Let $X\in\mathscr S_X \subset \mathbb R^d$ be a vector of payoff relevant variables, which are publicly observed by all players and also by the researcher.\footnote{See Section 5.4 on unobserved heterogeneity when this is not the case. See also \cite{grieco2011discrete} who analyzes a discrete game that has some payoff relevant variables publicly observed by all players, but not by the researcher. } For instance,  $X$ can include individual characteristics of the players as well as specific variables for the game environment.
For each player $i$, we further assume that the error term $U_i\in\mathbb R$ is her private information, i.e., $U_i$ is observed only by  player $i$, but not  by other players. To be consistent with the game theoretic literature, we also call $U_i$  the player $i$'s ``type'' \citep[see, e.g.,][]{fudenberg1991game}. Let  $U=(U_1,\cdots, U_{I})$ and $F_{U|X}$ be the conditional distribution function of $U$ given  $X$. The conditional  distribution $F_{U|X}$ is assumed to be common knowledge.

The payoff of player $i$ is described as follows:
\[
\Pi_i(Y,X,U_i)=\left\{\begin{array}{cc}\pi_i(Y_{-i},X)-U_i,&\text{ if } Y_i=1,\\0,&\text{ if } Y_i=0,\end{array}\right.
\] where $\pi_i$ is a structural function of interest.  The zero payoff for action $Y_i=0$ is a standard payoff normalization in binary response models.\footnote{Here we understand ``normalization'' from the view of observational equivalence: Suppose the payoffs take the  general form: 
\[
\Pi_i(Y,X,U^*_{i0},U^*_{i1})=\left\{\begin{array}{cc}\pi^*_{i1}(Y_{-i},X)-U^*_{i1},&\text{ if } Y_i=1;\\\pi^*_{i0}(Y_{-i},X)-U^*_{i0},&\text{ if } Y_i=0,\end{array}\right.
\] where for $y=0,1$, $U^*_{iy}$ and $\pi^*_{iy}$  are action--specific error terms and  payoff functions, respectively. It can be shown that this model with our subsequent assumptions is observationally equivalent to the above game with payoff  $\pi_i(Y_{-i},X)=\pi^*_{i1}(Y_{-i},X)-\pi^*_{i0}(Y_{-i},X)$ and $U_i=U^*_{i1}-U^*_{i0}$.}  

Following the literature on Bayesian games, given the public state variable $X$, player $i$'s decision rule is a function of her type:
\[
Y_i=\delta_i(X,U_i),
\]where $\delta_i:\mathbb{R}^{d}\times \mathbb{R}\rightarrow \{0,1\}$ maps all the information she knows to a binary decision.  For any given strategy profile $\delta=(\delta_1,\cdots,\delta_{I})$, let $\sigma^{\delta}_{-i}(a_{-i}|x,u_i)$ be  the conditional probability of other players choosing $a_{-i}\in\mathcal A_{-i}$ given $X=x$ and $U_i=u_i$, i.e., 
\[
\sigma^\delta_{-i}(a_{-i}|x,u_i)\equiv \Pr_{\delta}\left(Y_{-i}=a_{-i}|X=x,U_i=u_i\right)
 =\Pr\left[\delta_j(X, U_j)=a_{j}, \forall j\neq i|X=x,U_i=u_i\right],
 \]
where $\Pr_{\delta}$ represents the (conditional) probability measure under the strategy profile  $\delta$. 
The equilibrium concept we adopt is the pure strategy Bayesian Nash equilibrium (BNE). Mixed strategy equilibria are not considered in this paper, since a pure strategy BNE generally exists under  weak conditions in our model.   

We now characterize the equilibrium solution in our discrete game. Fix $X=x\in \mathscr S_X$. In equilibrium, player $i$ with $U_i=u_i$ chooses action $1$ if and only if her expected payoff is greater than zero, i.e., 
\begin{equation} \label{bestresponse}
\delta^*_{i}(x,u_i)=\textbf{1}\left[\sum_{a_{-i}}\pi_i(a_{-i},x)\times\sigma^*_{-i}(a_{-i}|x,u_i)-u_i\geq 0\right],\ \ \forall \ i,
\end{equation}
where $\delta^*\equiv(\delta^*_1,\cdots,\delta^*_{I})$, as a profile of functions of $u_1,\cdots,u_I$ respectively, denotes the equilibrium strategy profile and $\sigma^*_{-i}(a_{-i}|x,u_i)$ is a shorthand notation for $\sigma^{\delta^*}_{-i}(a_{-i}|x,u_i)$. Note that $\sigma^*_{-i}$ depends on $\delta^*_{-i}$. Hence, (\ref{bestresponse}) for $i=1,\cdots,I$ defines a simultaneous equation system in $\delta^*$ referred as ``mutual consistency'' of players' optimal behaviors.
A pure strategy BNE is a fixed point $\delta^*$ of such a system, which holds for all $u=(u_1,\ldots,u_I)$ in the support $\mathscr S_{U|X=x}$. Ensuring equilibrium existence in Bayesian games is a complex and deep subject in the literature. It is well known that a solution of such an equilibrium generally exists in a broad class of Bayesian games \citep[see, e.g., ][]{vives1990nash}.

The key to our approach is to employ a particular equilibrium solution concept of BNE --- monotone pure strategy BNEs, which  exist under additional weak conditions. Recently, much attention has focused on monotone pure strategy BNEs. The reason is that monotonicity is a natural property and has proven to be powerful in many applications such as auctions, entry, and global games. In our setting, a monotone pure strategy BNE is defined as follows:
\begin{definition}
Fix $x\in\mathscr S_X$. A pure strategy profile $\left(\delta_1^*(x,\cdot),\cdots,\delta_I^*(x,\cdot)\right)$ is a monotone pure strategy BNE if $\left(\delta_1^*(x,\cdot),\cdots,\delta_I^*(x,\cdot)\right)$ is a BNE and $\delta^*_i(x,u_i)$ is (weakly) monotone in $u_i$ for all $i$. 
\end{definition}

Monotone pure strategy BNEs are relatively easier to characterize than ordinary BNEs. Fix $X=x$. In our setting, a monotone pure strategy (m.p.s.) can be explicitly defined as a threshold  function (recall that $\delta_i^*$ can take only binary values).  Formally, in  an m.p.s. BNE, player $i$'s equilibrium strategy can be written as $\delta^*_i=\textbf{1} \left[u_i\leq u^*_i(x)\right]$,\footnote{The left--continuity of strategies considered hereafter is not restrictive given our assumptions below. Note that the payoff function is decreasing in $u_i$, hence the m.p.s. is also (weakly) decreasing. To simplify, throughout we use ``weakly/strictly monotone'' to refer to ``weakly/strictly  decreasing''.}  where $u^*_i(x)$ is the cutoff value that might depend on $x$. Let $u^*(x) \equiv (u^*_1(x),\cdots, u^*_I(x))\in \mathbb R^I$ be the profile of equilibrium strategy thresholds. 

In an m.p.s. BNE, the mutual consistency condition for a BNE solution defined by (\ref{bestresponse})  requires that  for each player $i$,
\begin{equation}
\label{multual_consistency}
u_i\leq u^*_i(x)\Longleftrightarrow \sum_{a_{-i}}\pi_i(a_{-i},x)\times\sigma^*_{-i}(a_{-i}|x,u_i)-u_i\geq 0.
\end{equation}  
A simple but key observation is that under certain weak conditions introduced later, (\ref{multual_consistency}) implies that player $i$ with the marginal type $u^*_i(x)$ should be indifferent between action $1$ and $0$, i.e.
\begin{equation}
\label{thresholds}
\mathbb \sum_{a_{-i}} \pi_i(a_{-i},x)\times\sigma^*_{-i}(a_{-i}|x,u^*_i(x))-u^*_i(x)=0.
\end{equation} Therefore, the equilibrium strategy can be represented by 
\begin{equation}
\label{representation_eq}
Y_i=\textbf 1\Big[U_i\leq \sum_{a_{-i}} \pi_i(a_{-i},X)\times\sigma^*_{-i}(a_{-i}|X,u^*_i(X))\Big].
\end{equation}


The seminal work on the existence of an m.p.s. BNE in games of incomplete information was first provided by \cite{athey2001single} in both \textit {supermodular} and \textit{logsupermodular} games, and later extended by \cite{mcadams2003isotone} and  \cite{reny2011existence}. Applying \cite{reny2011existence} Theorem 4.1, we establish the existence of m.p.s.  BNEs in our binary game under some weak regularity assumptions.
\begin{CRND}[Conditional Radon--Nikodym Density]
\label{continuity}
For every $x \in \mathscr S_X$, the conditional distribution of $U$ given $X=x$ is absolutely continuous w.r.t.  Lebesgue measure and has a   continuous positive conditional Radon--Nikodym density $f_{U|X}(\cdot|x)$ a.e. over the nonempty interior of its hypercube support $\mathscr S_{U|X=x}$.
\end{CRND}
\noindent
Assumption R allows the support of $U$ conditional on $X=x$ to be bounded, namely of the form $\times_{i=1,\dots, I} [\underline{u}_i(x), \overline{u}_i(x)]$ for some finite endpoints $\underline{u}_i(x)$ and $\overline{u}_i(x)$ as frequently used when $U_i$ is $i$'s private information, or unbounded such as when  $\mathscr S_{U|X=x}=\mathbb R^I$  in binary response models.  As a matter of fact, assumption R can be greatly weakened as shown by \cite{reny2011existence} (see \Cref{rationalizing_all} for more details).

For any strategy profile $\delta$, let $\mathbb E _{\delta}$ denote the (conditional) expectation under the strategy profile $\delta$. Without causing any confusion, we will suppress the subscript $\delta^*$ in $\mathbb E _{\delta^*}$ (or $\Pr_{\delta^*}$) when the expectation (or probability) is taken under the equilibrium strategy profile.

\begin{MEP}[Monotone Expected Payoff]
For any weakly m.p.s. profile $\delta$ and $x\in\mathscr S_X$, the (conditional) expected payoff $\mathbb E _\delta \left[\pi_i(Y_{-i},X)\big| X=x,U_i=u_i\right]-u_i$ is a  weakly monotone function in $u_i \in \mathscr S_{U_i|X=x}$.\end{MEP} 
\noindent
Assumption M guarantees that each player's best response is also weakly monotone in type given that all other players adopt weakly m.p.s.. In particular, in a two--player game (i.e., $I=2$), assumption M is equivalent to the following condition: for any $x\in\mathscr S_X$ and $u_{-i}\in\mathbb R$, the function $\left[\pi_i(1,x)-\pi_i(0,x)\right]\times \Pr (U_{-i}\leq {u_{-i}}|X=x,U_i=u_i)- u_i$ is weakly monotone in $u_i$.

 Note that  for the existence of m.p.s. BNEs, assumption M is sufficient but not necessary in many cases. It should also be noted that assumption M holds trivially if all $U_i$s are conditionally independent of each other given $X$. \Cref{msbe1} in \Cref{existence_msbe} also provides primitive sufficient conditions for assumption M.  Specifically, we assume positive regression dependence across $U_i$s given $X$ and strategic complementarity of players' actions, which are natural restrictions in models with social interactions. 
\begin{lemma}
\label{lemma1}
Suppose assumptions R and M hold.  For any $x\in\mathscr S_X$, there exists an m.p.s. BNE.  
In particular, player $i$'s equilibrium strategy can be written as in \eqref{representation_eq}. 
\proof See \Cref{proof_lemma1}
\qed
\end{lemma}

By \Cref{lemma1}, m.p.s. BNEs generally exist in a large class of binary games.  As far as we know, with the only exception of \cite{aradillas2008identification} and \cite{xu2010estimation}, every paper analyzing empirical discrete games of incomplete information so far has imposed certain restrictions (i.e. sufficient conditions for assumption M) to guarantee that equilibrium strategies be threshold--crossing.

Monotone pure strategy BNEs are  convenient and powerful for empirical analysis. In particular, we can represent each player's equilibrium strategy by a semi--linear--index binary response model (\ref{representation_eq}). Such a representation relates to single-agent binary threshold crossing models studied by e.g. \cite{matzkin1992nonparametric}, where the `coefficients' are the player's equilibrium belief about the other players' actions evaluated at the player's equilibrium signal threshold $u_i^*(x)$.  Note, however, that we do not restrict either $\pi_i(a_{-i},X)$ or $F_{U|X}$ to have a specific functional form.  Nevertheless, in Section 4.2 we will show that the equilibrium beliefs  $\sigma^*_{-i}$ in (\ref{representation_eq}) can be nonparametrically identified under additional weak conditions.  

Though non--monotone strategy BNEs are seldom considered in the literature, it is worth pointing out that this kind of equilibria could exist and sometimes even stands as the only type of equilibria. This could happen when some player is quite sensitive to others' choices and types are highly correlated.   We provide a simple example to illustrate.\footnote{We thank Steven Stern and Elie Tamer for their comments and suggestions on the following example.}


\begin{example}
\label{example1}
Let $I=2$ and  $\pi_i=X_i-\beta_i Y_{-i}  - U_i$, where  $(U_1,U_2)$ conforms to a joint  normal distribution with mean zero, unit variances and correlation parameter $\rho\in(-1,1)$.\footnote{In a similar fully parametric setting, \cite{xu2010estimation} proposes an inference approach based on   first identifying a subset of the covariate space where  the game admits a unique m.p.s. BNE.} 

Case 1: Suppose $(X_1,X_2)=(1,0)$ and $(\beta_1,\beta_2)=(2, 0)$. Then, regardless of the value of $\rho$, there is always a unique pure strategy BNE: Clearly, player 2 has a dominant strategy which is monotone in $u_2$: choosing 1 if and only if $u_2\leq 0$. Thus, player 1's best response must be: choosing 1 if and only if $1-2 \Phi\big(-\frac{\rho u_1}{\sqrt{1-\rho^2}}\big)-u_1\geq 0$. Further, it can be shown that player 1's equilibrium strategy is  not monotone in $u_1$ if and only if $\rho\in\big( \sqrt{\frac{\pi}{2+\pi}}, 1) $. 

Case 2:  Suppose $(X_1,X_2)=(1,1)$ and $(\beta_1,\beta_2)=(2, 2)$. First, note that the m.p.s. profile $\{1(u_1\leq 0);1(u_2\leq 0)\}$ is  a BNE as long as $\rho\in\big(-1, \sqrt{\frac{\pi}{2+\pi}} \ \big] $. Moreover, it can be verified that this equilibrium is the unique BNE if and only if $\rho\in\big(-1, \frac{\pi-2}{2+\pi} \ \big] $. When $\rho\in\big( \frac{\pi-2}{2+\pi}, 1 \ \big) $, we can find two other equilibria of the game: $\{1(u_1\leq u^*);1(u_2\leq -u^*)\}$ and $\{1(u_1\leq -u^*);1(u_2\leq u^*)\}$ where $u^*>0$ solves $
1-2\Phi(\sqrt{\frac{1+\rho}{1-\rho}}\cdot u^*) +u^*=0$.

\end{example}
In Case 1, the non--monotone strategy BNE occurs due to the large positive correlation between $U_1$ and $U_2$ (relative to $\beta_1$), which violates assumption M.  On the other hand, for any given  value of the structural parameters, the existence of a non--monotone strategy BNE could also depend on the realization of $(X_1,X_2)$.

\section{Rationalization}
\label{sec:rationalization}
In this section, we study the baseline model defined by assumptions R and M as well as two other models obtained by imposing additional assumptions frequently made in the empirical game literature.  Specifically,  we  characterize all the restrictions imposed on the distribution of observables $(Y,X)$ by each of these models. 

We say that a conditional distribution  $F_{Y|X}$ is rationalized by a model if and only if it satisfies all the restrictions of the model.  Equivalently, $F_{Y|X}$ is rationalized by the model if and only if there is a structure (not necessarily unique) in the model that generates such a distribution. In particular, rationalization logically precedes identification as the latter, which is addressed in Section 4, makes sense only if the observed distribution can be rationalized by the  model under consideration.

An assumption frequently made in the literature, e.g., \cite{brock2001discrete} and \cite{bajari2010estimating}, is the exogeneity of the observed state variables $X$ relative to private information $U$.  To our knowledge, exceptions are \cite{de2012inference} and \cite{wan2010semiparametric}.
\begin{E}[Exogeneity]
$X$ and $U$ are independent of each other.\footnote{\label{footnote9}Our results can be easily extended to the weaker assumption that  $X$ and $U$ are independent from each other conditional on $W$, where $W$ are other observed payoff relevant variables.}
\end{E}

Another assumption called as mutual independence is also widely used in the literature. For examples, see an extensive list of references in  two recent surveys: \cite{bajari2010game} and \cite{de2012econometric}. Such an independence of types is a convenient theoretical assumption, which means player $i$'s private information is uninformative about other players' types given $X$. 
\begin{MI}[Mutual Independence]
$U_1,\cdots,U_I$ are mutually independent conditional on $X$.
\end{MI}

Let $S\equiv [\pi; F_{U|X}]$, where $\pi=(\pi_1,\cdots,\pi_I)$.  We now consider the following models: 
\begin{align*}
&\mathcal M_1\equiv\big\{S:  \text{Assumptions R and M hold and a single m.p.s. BNE is played}\big\},\\
&\mathcal M_2\equiv\left\{ S \in\mathcal M_1:  \text{ Assumption E holds}\right\},\\
&\mathcal M_3 \equiv\left \{ S\in \mathcal M_2 : \ \text{Assumption I holds}  \right\}.
\end{align*}
Clearly, $\mathcal M_1\supsetneq\mathcal M_2\supsetneq\mathcal M_3$. 

The last requirement in $\mathcal M_1$ is not restrictive when  the game has a unique equilibrium which has to be an m.p.s. BNE. In the global game literature, for example, one achieves uniqueness of m.p.s. BNE as the information noise gets small. See e.g. \cite{carlsson1993global} and \cite{morris1998unique}. In social interactions, uniqueness of  m.p.s. BNE has also been established in e.g. \cite{brock2001discrete} and \cite{xu2011social}. When there exist multiple equilibria, we follow part of the literature by assuming that the same equilibrium is played in the DGP for any given $x$.  See e.g. \cite{aguirregabiria2013recent} for a survey. Such an assumption is realistic if the equilibrium selection rule is actually governed by some game invariant factors, like culture, social norm, etc. See, e.g.,  \cite{de2012econometric} for a detailed discussion.  Relaxing such a requirement has been addressed in recent work and will be discussed in Section 5.3.

We introduce some key notation for the following analysis. For any structure $S\in\mathcal M_1$, let $\alpha_i(x)\equiv F_{U_i|X}(u^*_i(x)|x)$.  By monotonicity of the equilibrium strategy, we have $\alpha_i(x)=\mathbb E(Y_i|X=x)$, i.e. $\alpha_i(x)$ is player $i$'s (marginal) probability of choosing action $1$ given $X=x$.  Moreover, for each $p=2,\cdots,I$, and $1\leq i_1<\cdots<i_p\leq I$, let $ C_{U_{i_1},\cdots,U_{i_p}|X}$ be the conditional copula function of $(U_{i_1},\cdots, U_{i_p})$ given $X$, i.e.,  for any $(\alpha_{i_1},\cdots,\alpha_{i_p})\in [0,1]^p$ and $x\in\mathscr S_X$,
\[
 C_{U_{i_1},\cdots,U_{i_p}|X}( \alpha_{i_1},\cdots,\alpha_{i_p}|x)\equiv F_{U_{i_1},\cdots,U_{i_p}|X}\left(F^{-1}_{U_{i_1|X}}(\alpha_{i_1}|x),\cdots, F^{-1}_{U_{i_p|X}}(\alpha_{i_p}|x)\Big| x\right).
\]

The next proposition characterizes the collection of distributions of $Y$ given $X$ that can be {\em rationalized} by $\mathcal M_1$.
\begin{proposition}
\label{lemma_oes_1}
A conditional distribution $F_{Y|X}$ is rationalized by $\mathcal M_1$ if and only if for all $x\in\mathscr S_X$ and $a\in\mathcal A$,
 $\Pr(Y=a|X=x)=0$ implies that $\Pr(Y_i=a_i|X=x)=0$ for some $i$.
\proof
See \Cref{proof_lemma_oes_1}
\qed
\end{proposition}
By \Cref{lemma_oes_1}, $\mathcal M_1$ rationalizes all distributions for $Y$ given $X$ that belong to the interior of the  $2^{I}-1$ dimensional simplex, i.e. distributions with strictly positive choice probabilities, since the condition  in \Cref{lemma_oes_1} is void for such distributions.  Specifically, the distributions that cannot be rationalized by $\mathcal M_1$ must have $\Pr(Y=a|X=x)=0$ for some $a\in\mathcal A$, i.e., distributions for which there are ``structural zeros.''  In other words, our baseline model $\mathcal M_1$ imposes  no essential restrictions on the distribution of observables. 
The distributions that cannot be rationalized by  $\mathcal M_1$  arise because of assumption R.  As noted earlier, one can replace assumption R by \cite{reny2011existence}'s weaker conditions, in which case any distribution for $Y$ given $X$ can be rationalized.   See \Cref{lemma_oes_2} in \Cref{rationalizing_all}.

We now characterize all the restrictions imposed on $F_{Y|X}$ by model $\mathcal M_2$.  These additional restrictions come from assumption E.
\begin{proposition}
\label{coro1}
A conditional distribution $F_{Y|X}$ rationalized by  $\mathcal{M}_1$ is also rationalized by   $\mathcal{M}_2$ if and only if  for each $p=2,\cdots,I$ and $ 1\leq i_1<\cdots<i_p\leq I$,
\begin{itemize}
\setlength{\itemsep}{6pt}
\item [R1:]  
$
  \mathbb E \big(\prod_{j=1}^pY_{i_j}\big|X\big)
= \mathbb E \big(\prod_{j=1}^pY_{i_j}\big| \alpha_{i_1}(X), \cdots, \alpha_{i_p}(X)\big).
$

\item[R2:] $\mathbb E\big(\prod_{j=1}^pY_{i_j}|\alpha_{i_1}(X)=\cdot, \cdots, \alpha_{i_p}(X)=\cdot\big)$ is strictly  increasing on $\mathscr S_{\alpha_{i_1}(X),\cdots,\alpha_{i_p}(X)}$ except at values
for which some coordinates are zero.

\item[R3:]  $\mathbb E\big(\prod_{j=1}^pY_{i_j}|\alpha_{i_1}(X)=\cdot, \cdots, \alpha_{i_p}(X)=\cdot\big)$ is continuously differentiable  on $\mathscr S_{\alpha_{i_1}(X),\cdots,\alpha _{i_p}(X)}$.
 \end{itemize}
\proof
See \Cref{proof_coro1}.
\qed
\end{proposition}
In \Cref{coro1}, the most stringent restriction  is R1, which requires that the joint choice probability depend on $X$ only through the corresponding marginal choice probabilities.  Under restrictions R1 and R2, the condition $\alpha(x)\geq \alpha(x')$ implies that $\Pr (Y_{i_1}=1,\cdots,Y_{i_p}=1|X=x)\geq \Pr (Y_{i_1}=1,\cdots,Y_{i_p}=1|X=x')$ for all tuples $ \{ i_1,\cdots,i_p\}$. Moreover, note that $\alpha_i(x)$ is identified by  $\alpha_i(x)=\mathbb E (Y_i|X=x)$.  Therefore, all the restrictions R1--R3 are testable in principle.\footnote{ If $X$ is discrete, then R3 becomes irrelevant.}  This is discussed further in the Conclusion.

For completeness, we also study the restrictions on  observables imposed by $\mathcal M_3$, which makes the additional assumption I. It should be noted that assumption M is satisfied when assumption I holds.  In other words,  
\[
\mathcal M_3=\{S: \text{Assumptions R, E, I hold and a single m.p.s. BNE is played}\}.
\] 
In the literature, several special cases of $\mathcal M_3$ have been considered under some parametric or functional form restrictions, see, e.g., 
 \cite{bajari2010estimating} and \cite{lewbel2012identification}.
 \begin{proposition}
\label{ration_M_4}
A conditional distribution $F_{Y|X}$ can be rationalized by   $\mathcal{M}_3$ if and only if $Y_1,\cdots,Y_I$ are conditionally independent given $X$, i.e. for each $p=2,\cdots,I$ and $ 1\leq i_1<\cdots<i_p\leq I$, $\mathbb E\big(\prod_{j=1}^pY_{i_j}|X\big)=\prod_{j=1}^p\alpha_{i_j}(X)$.
\proof
See \Cref{proof_ration_M_4}.
\qed
\end{proposition}

It is worth pointing out that \Cref{lemma_oes_1,coro1,ration_M_4} exhaust all possible testable restrictions as they provide necessary {\em and} sufficient conditions for rationalizing $\mathcal{M}_1$, $\mathcal{M}_2$ and $\mathcal{M}_3$, respectively.   Moreover, their proofs are constructive. Specifically, we construct an $I$--single--agent decision structure that rationalizes the given distributions satisfying the corresponding restrictions. This is summarized by the following corollary. For $k=1,2,3$, let  
\[
\mathcal M^s_k=\{S\in \mathcal M_k : \ \pi_i(a'_{-i},x)=\pi_i(a_{-i},x), \forall a'_{-i}, a_{-i}\in\mathcal A_{-i}, x\in\mathscr S_X \text{ and } i=1,\cdots,I \}.
\] 
\begin{corollary}
\label{corolary0}
For $k=1,2,3$, $\mathcal M_k$ is observationally equivalent to $\mathcal M^s_k$.
\end{corollary} 
\noindent
Hence, it is evident that without additional model restrictions beyond the assumptions of $\mathcal M_1$ $\mathcal M_2$, or $\mathcal M_3$, a discrete Bayesian game model with strict interactions cannot be empirically distinguished from an alternative model with $I$--single--agent decisions.   In contrast, with exclusion restrictions, Section 4 shows that these two classes of models can be distinguished from each other.

It should also be noted that  the conditional independence restriction in \Cref{ration_M_4} implies conditions R1--R3 in \Cref{coro1}, as well as the necessary and sufficient condition in \Cref{lemma_oes_1}.  The conditional independence of players' choices given payoffs shifters characterizing  $\mathcal M_3$ suggests that we can replace $\mathcal M_3$ in \Cref{ration_M_4} with  
\[
\mathcal M_3'\equiv
\{S: \text{ Assumption I holds and a single BNE is played}\},
\]
since $\mathcal M'_3$ also implies the conditional independence restriction. In particular, $\mathcal M'_3$ does not require the monotonicity of the BNE. Because $\mathcal M_3\subset \mathcal M_3' $, we have the following corollary.
\begin{corollary}
\label{corolary1}
Model $\mathcal M_3$ imposes the same restrictions on the distribution
of observables as $\mathcal M_3'$, i.e., both models are observationally equivalent.
\end{corollary}
\noindent
This is a surprising result: Assumptions R, M and more importantly exogeneity of the payoff shifters (assumption E) become redundant in terms of 
restrictions on the observables, as soon as mutual independence of types conditional on $X$ (assumption I) and a single BNE condition are imposed on the baseline model.  Moreover, if we are willing to maintain assumption I, then \Cref{ration_M_4} and \Cref{corolary1} gives us a test of a single equilibrium being played, as rejecting the conditional independence of the players' choices given $X$ indicates the presence of multiple equilibria. This extends a related result in terms of correlation obtained by \cite{de2012inference} in a partial--linear setting.

\section{Nonparametric Identification} \label{section id}

In this section we study the nonparametric identification of
the baseline game-theoretic model $\mathcal M_1$, and its special cases $\mathcal M_2$ and $\mathcal M_3$. The recent literature has focused on the parametric or semiparametric identification of structures in $\mathcal M_3$, see, e.g.,      \cite{brock2001discrete,seim2006empirical,sweeting2009strategic,bajari2010estimating}, and \cite{tang2010estimating}. As far as we know,  \cite{lewbel2012identification}  is the only paper that studies the nonparametric identification of  a submodel of $\mathcal M_3$ obtained through additional restrictions on the functional form of payoffs.

In our context, identification of  each model is equivalent to identification of the payoffs  $\pi_i$, the marginal distribution function $F_{U_i|X}$ and the copula function $C_{U|X}$ of the joint distribution of private information. Let $Q_{U_i|X}$ be the quantile function of $F_{U_i|X}$. Because the quantile function is the inverse of the CDF, i.e. $Q_{U_i|X}= F^{-1}_{U_i|X}$,  identification reduces to that of the triple $\left[\pi; \{Q_{U_i|X}\}_{i=1}^I;C_{U|X}\right]$.  In contrast to single-agent binary threshold crossing models, we do not require any of such primitives to be parameterized.


We first show that $\mathcal M_1$ is not identified in general. For $\mathcal M_2$, we first establish the identification of $C_{U|X}$ and the equilibrium beliefs $\sigma^*_{-i}(\cdot|x,u^*_i(x))$ under an additional support condition. The identification of the copula function $C_{U|X}$ is of particular interest in applications on social interactions, since it represents homophily among friends. Under  some exclusion restrictions and rank conditions, we then establish the full identification of the payoff functions $\pi$ and the  quantile functions $ \{Q_{U_i|X}\}_{i=1}^I$  up to a single location--and--scale normalization on the payoffs.  Regarding $\mathcal M_3$, its identification requires slightly weaker support restrictions than those for $\mathcal M_2$, though the differences  are not essential.

\subsection{ Nonidentification of $\mathcal M_1$  }
We begin with the most general model $\mathcal M_1$. 
\begin{proposition}
$\mathcal M_1$ is not identified nonparametrically.
\end{proposition}
\noindent
The proof is trivial. It follows directly from the observational equivalence between any structure $S$ in $\mathcal M_1$ and a collection of $I$--single-agent binary responses models: Let $\tilde S\equiv \left(\tilde\pi; \tilde F_{U|X}\right)$, in which $\tilde \pi_i(\cdot,x)\equiv\tilde u_i(x)$, where $\tilde u_i(x)$ is arbitrarily chosen, and   $\tilde F_{U|X}$ satisfies assumption $R$ with $\tilde F_{U_{i_1},\cdots,U_{i_p}|X}(\tilde u_{i_1}(x), \cdots, \tilde u_{i_p}(x)|x)=F_{U_{i_1},\cdots, U_{i_p}|X}(u_{i_1}^*(x),\cdots, u_{i_p}^*(x)|x)$ for all $x\in\mathscr S_X$ and all tuples $ \{ i_1,\cdots,i_p\}$. Thus, $\tilde S$ and $S$ are observationally equivalent thereby establishing the non--identification of $\mathcal M_1$.\footnote{Even if we impose the identifying restrictions (namely the exclusion restriction, support condition and  rank condition)  introduced later, $\mathcal M_1$ is still not identified by a similar argument. }

Next, we turn to the identification of  $\mathcal M_2$ and its sub--model $\mathcal M_3$. First note that we maintain assumption E in both models, i.e. that $X$ and $U$ are independent of each other; see \cref{footnote9} for a weaker assumption. It follows that $Q_{U_i|X}=Q_{U_i}$ and  $C_{U|X}=C_U$.  Thus, identification of these models reduces to that of the triple $\left[\pi; \{Q_{U_i}\}_{i=1}^I; C_U\right]$.

\subsection{Identification of $\mathcal M_2$}
Let $\alpha(x)\equiv\left(\alpha_1(x),\cdots,\alpha_I(x)\right)$ be a profile of the marginal choice probabilities. Note that $\alpha_i(x)$ is identified by $\mathbb E (Y_i|X=x)$.  Under assumption E, the copula function $C_U$ is nonparametrically identified 
on an appropriate domain, namely, the extended support of $\alpha(X)$ defined as $\mathscr S^e_{\alpha(X)}\equiv \{\alpha: \alpha_j=0 \text{ for some }j\}\bigcup \big\{\alpha: (\alpha_{i_1}, \cdots,  \alpha_{i_p})\in\mathscr S_{\alpha_{i_1}(X),\cdots, \alpha_{i_p}(X)}; \text{other }\alpha_{i_j}=1\big\}$. We have $C_U(\alpha)=0$ if $\alpha_j=0$ for some $j$; otherwise,\footnote{In (\ref{iden_copula}), it is understood that $C_{U}(\alpha)=1$ if $\alpha=(1,\cdots,1)$.} 
\begin{multline}
\label{iden_copula}
C_{U}(\alpha)
=\Pr\big\{ U_{1}\leq Q_{U_1}(\alpha_{1}),\cdots,U_{I}\leq Q_{U_I}(\alpha_{I}) \big\}\\
=\Pr\left\{ U_{j}\leq Q_{U_j}(\alpha_{j}), \ \forall j\in \{i: \alpha_i\neq 1\}\right\}
=\mathbb E \Big\{\prod_{i=1}^IY_{i}\big|\alpha_{j}(X)=\alpha_{j},  \forall j\in \{i: \alpha_i\neq 1\}\Big\}.
\end{multline}
Key among those conditions for the nonparametric identification of $C_U$ is the assumption that a single m.p.s. BNE  is played in the DGP. Such a restriction implies that conditional on $\alpha_j(X)=\alpha_j$, the event $U_j\leq Q_{U_j}(\alpha_{j})$ is equivalent to $Y_j=1$.

As mentioned above, the equilibrium belief $\sigma^*_{-i}(\cdot|x,u^*_i(x))$ can also be nonparametrically identified, for which we need a support condition on ${\alpha(X)}$. 
\begin{SC}[Support Condition]
\label{diffz}
The support $\mathscr S_{\alpha(X)}$ is a convex subset of $[0,1]^I$ with full dimension, i.e. $\text{dim}\big(\mathscr S_{\alpha(X)}\big)=I$.
\end{SC}

\noindent
Assumption SC implies that the relative interior and the interior $\mathscr S^\circ_{\alpha(X)}$ of $\mathscr S_{\alpha(X)}$ are equal. Hence, the dimension of the interior of $\mathscr S_{\alpha(X)}$ is $I$. See e.g. \cite{rockafellar1997convex}. Therefore, we can take derivatives in all directions of an arbitrary smooth function defined on $\mathscr S^\circ_{\alpha(X)}$.\footnote{As a matter of fact, for the boundary points, we can take directional derivatives as well.} Moreover, given the identification of $\alpha(x)$, assumption SC is verifiable.

Assumption SC is high level, requiring that the payoff shifters $X$ contain at least one continuously distributed component. For instance, suppose  $\pi_i(a_{-i},x)=\pi_{i0}(x)+\beta_{i}\times  \sum_{j\neq i}a_{j} $ where $\pi_{i0}(\cdot):\mathscr S_X\rightarrow \mathbb R$, $\beta_i\in\mathbb R^+$ and $(U_1,\cdots, U_I)$ are positively regression dependent. Given assumption E, \eqref{thresholds} becomes
\begin{equation}
\label{x_res}
\pi_{i0}(x) =u^*_i(x) -\beta_i\times \sum_{j\neq i} \mathbb P(U_j\leq u_j^*(x)|U_i=u_i^*(x)), \ \ \ \forall i=1,\cdots,I.
\end{equation}Because $u_i^*(x)=Q_{U_i}(\alpha_i(x))$,  the support of $(\pi_{10}(X),\cdots,\pi_{I0}(X))$  can be derived from the support $\mathscr S_{\alpha(x)}$ and the joint distribution of $(U_1,\cdots, U_I)$. In general, the support condition SC requires $(\pi_{10}(X),\cdots,\pi_{I0}(X))$ to be continuously distributed on a support with full dimension.
Moreover, the full dimensionality of  $\mathscr S_{\alpha(X)}$ requires neither exclusion restrictions, nor the dimension of $X$ to be larger than or equal to the number of players.  To see this, consider the extreme situation where $\mathscr S_{\alpha(X)}$ has full support $[0,1]^I$. In the above example,  suppose  the dimension of $X$ is one and $\mathscr S_X= \mathbb R$. Let $(\pi_{10}(\cdot),\cdots,\pi_{I0}(\cdot)):\mathbb R\rightarrow \mathbb R^I$ be a one--to--one and onto mapping.\footnote{Following the Cantor--Schroeder--Bernstein Theorem, one can explicitly construct such a mapping. The constructed mapping is not homeomorphic due to discontinuity.} It follows that for any $\alpha\in[0,1]^I$, one can find a value $x\in \mathbb R$ that satisfies \eqref{x_res} and could induce $\alpha(x)=\alpha$.  On the other hand, suppose the payoffs satisfy the linear--index specification  $\pi_{i0}(x)=x'\theta_i$ where $x\in\mathbb R^k$. Then, the full support of $\mathscr S_{\alpha(X)}$ requires that the index profile $(X'\theta_1,\cdots,X'\theta_I)$ has full support $\mathbb R^I$. This holds, for instance, when $X$ contains an $I$--dimensional  sub--vector $X_s$ with full support on $\mathbb R^I$ conditional on $X_{-s}$, and the corresponding square submatrix $(\theta_{s1},\cdots, \theta_{sI})$ has full rank.  A similar condition has been considered in \cite{lewbel2012identification} using special regressors.

\begin{lemma}
\label{iden0}
Let $S\in\mathcal M_2$.  Suppose assumption SC holds. Fix $x\in\mathscr S^\circ_X$.  Then the equilibrium beliefs $\sigma^*_{-i}(\cdot|x,u^*_i(x))$ are identified, namely, for all $a_{-i}\in\mathcal A_{-i}$,
\begin{equation}
\label{belief_eq}
\sigma^*_{-i}(a_{-i}|x,u^*_i(x))
=\frac{\partial \Pr \left(Y_i=1; Y_{-i}=a_{-i}|\alpha(X)=\alpha \right)}{\partial \alpha_i}\Bigg|_{\alpha=\alpha(x)}.
\end{equation}
\proof See \Cref{app2}.
\qed
\end{lemma}

Note that, under assumption I, the probability $\Pr \left(Y_i=1; Y_{-i}=a_{-i}|\alpha(X)\right)=\alpha_i(X)\times\prod_{j\neq i}\alpha^{a_j}_j(X)[1-\alpha_j(X)]^{1-a_j}$ becomes a (known) linear function in $\alpha_{i}(X)$. Thus, we have $\sigma^*_{-i}(a_{-i}|x,u^*_i(x))=\prod_{j\neq i}\alpha^{a_j}_j(X)[1-\alpha_j(X)]^{1-a_j}$, thereby identifying trivially the equilibrium beliefs without assumption SC, see, e.g., \cite{bajari2010estimating}.

The intuition for our identification of $\sigma^*_{-i}$ is acquired from the local--instrumental--variables method developed in \cite{heckman1999local,heckman2005marginal}: The local variation in player $i$'s choice probability $\alpha_i(X)$ (after controlling for $\alpha_{-i}(X)$) provides the identification power for the conditional choice probability of the other players given that the latent variable is at the margin. To illustrate this, we use a two--player game. 
\begin{example}
\label{example}
Let $S\in\mathcal M_2$ and $I=2$.  Note that for any $\alpha\in [0,1]^2$, we have 
\[
\Pr(Y_{1}=1,Y_2=1|\alpha(X)=\alpha)
=\Pr\left[U_1\leq Q_{U_1}(\alpha_1), U_2\leq Q_{U_2}(\alpha_2)\right]
=C_{U}(\alpha_1, \alpha_2),
\] where the first equality follows from $\alpha_i(X)=\alpha_i$ being equivalent to $u^*_i(X)=Q_{U_i}(\alpha_i)$ from the independence of $U$ and $X$. Further, we have 
\begin{equation}
\label{copula_diff}
\frac{\partial C_U(\alpha_1, \alpha_2)}{\partial \alpha_i}=\Pr\left(U_{-i}\leq Q_{U_{-i}}(\alpha_{-i})|U_i=Q_{U_i}(\alpha_i)\right),
\end{equation} see, e.g., \cite{darsow1992copulas}. Because $Q_{U_{i}}(\alpha_{i}(x))=u_i^*(x)$, it follows that
\begin{multline*}
\frac{\partial \Pr(Y_{1}=1,Y_2=1|\alpha(X)=\alpha)}{\partial \alpha_i}\Big|_{\alpha=\alpha(x)}=\Pr\left(U_{-i}\leq u^*_{-i}(x)|U_i=u^*_i(x)\right)\\
=\Pr\left(Y_{-i}=1|X=x, U_i=u^*_i(x)\right)=\sigma^*_{-i}(1|x, u^*_i(x)).
\end{multline*}
Similarly, we have  
\[
\frac{\partial \Pr(Y_i=1, Y_{-i}=0|\alpha(X)=\alpha)}{\partial \alpha_i}\Big|_{\alpha=\alpha(x)}=\sigma^*_{-i}(0|x, u^*_i(x)).
\]

\end{example}

Equation (\ref{copula_diff}) is related to the treatment effect literature, e.g., \cite{heckman1999local,heckman2005marginal}, \cite{carneiro2009estimating} and \cite{jun2011tighter}. Taking derivative with respect to the propensity score identifies the conditional quantile (or conditional expectation)  of the treatment effect at the margin. \Cref{iden0} extends this result to the multivariate case by using the law of iterated expectation.

We now discuss the identification of $\pi$ and $Q_{U_i}$.  Fix $X=x$ such that $\alpha_i(x)\in (0,1)$.  By the proof of \Cref{lemma1} and the fact that $u^*_i(x)=Q_{U_i}(\alpha_i(x))$, we can represent the equilibrium condition (\ref{thresholds}) by 
\begin{equation}
\label{eqcond}
\sum_{a_{-i}\in\mathcal A_{-i}}\pi_i(a_{-i},x)\times \sigma^*_{-i}(a_{-i}|x,u^*_i(x))-Q_{U_i}(\alpha_i(x))=0,
\end{equation} in which $\sigma^*_{-i}$ is known by  \Cref{iden0}.  Next, we will exploit (\ref{eqcond}) for the identification of the payoffs $\pi_i$. The idea is to vary $\sigma^*_{-i}(\cdot|x,u^*_i(x))$ while keeping $\pi_i(\cdot, x)$ fixed, for which we need the following exclusion restriction. 
\begin{ER}[Exclusion Restriction] \label{exclusion}
Let $X = (X_1,\cdots, X_{I})$. For all $i$, $a_{-i}$ and $x$, we have $\pi_i(a_{-i}, x)=\pi_i(a_{-i},x_i)$.\footnote{As a matter of fact, $X_i$s can have some common variables due to homophily. In this case, our results hold by conditioning on those common variables.}
\end{ER}
\noindent
In the context of discrete games, the identification power of exclusion restrictions  was first demonstrated in \cite{pesendorfer2003identification}, \cite{tamer2003incomplete}, and  was used by \cite{bajari2010estimating} in a semiparametric setting. For instance,  in empirical IO,  some cost shifters  are included in the payoff of firm $i$ but not in firm $j$'s, and vice versa.

Under assumption ER, (\ref{eqcond}) implies that
\[
\sum_{a_{-i}\in\mathcal{A}_{-i}} \pi_i(a_{-i},x_i)\cdot\Big\{ \sigma^*_{-i}(a_{-i}|x,u^*_i(x))-\mathbb E \left[\sigma^*_{-i}\big(a_{-i}|X,u^*_i(X))\big)|X_i=x_i,\alpha_i(X)=\alpha_i(x)\right]\Big\}
=0.
\]
For notational simplicity, we  denote the random vector $\sigma^*_{-i}\left(\cdot|X,u^*_i(X)\right)$ as $\Sigma^*_{-i}(X)$, a column vector of dimension $2^{I-1}$.   Let $\overline \Sigma^*_{-i}(X)\equiv\Sigma^*_{-i}(X)-\mathbb E \left[\Sigma^*_{-i}(X)|X_{i},\alpha_i(X)\right]$ and ${\mathcal R}_i(x_i)=\mathbb E  \left[\overline\Sigma^*_{-i}(X){\overline\Sigma^*_{-i}}(X)^\top\big|X_i=x_i\right]$. Given \Cref{iden0}, we treat $\Sigma^*_{-i}(X)$ and $\overline \Sigma^*_{-i}(X)$ as observables hereafter. Note that  $\iota' \Sigma^*_{-i}(X)=1$ a.s., where $\iota\equiv(1,\cdots,1)'\in \mathbb R^{2^{I-1}}$. It follows that $\iota'\overline \Sigma^*_{-i}(X)=0$. Thus, $\overline \Sigma^*_{-i}(X)$ consists of a vector of linearly dependent variables. Indeed, the largest possible rank of the matrix ${\mathcal R}_i(x_i)$ is $2^{I-1}-1$, which implies there are no strategic interactions in player $i$'s decision at equilibrium. In the next proposition, we give identification results for features of $\mathcal M_2$.

\begin{lemma}
\label{idco}
Suppose $S\in\mathcal M_2$, and assumptions SC and ER hold.  Fix $x_i\in\mathscr S_{X_i}$ such that $\mathscr S_{\alpha_i(X)|X_i=x_i}\bigcap (0,1)\neq \emptyset$.  If the rank of ${\mathcal R}_i(x_i)$ is $2^{I-1}-1$, then $\mathscr S_{\alpha_i(X)|X_i=x_i}$ must be a singleton $\{\alpha^\dag_i\}$ and $\pi_i(\cdot,x_i)$ is identified up to the $\alpha^\dag_i$--quantile of $F_{U_i}$, i.e. $\pi_i(\cdot,x_i)=Q_{U_i}(\alpha^\dag_i)$. If the rank of ${\mathcal R}_i(x_i)$ is $2^{I-1}-2$, then $\pi_i(\cdot,x_i)$ is identified up to location and scale that depend on $x_i$, or equivalently, $\pi_i(\cdot,x_i)-\pi_i(a^0_{-i},x_i)$ is identified up to scale for arbitrary $a^0_{-i}\in\mathcal A_{-i}$. 
\proof
See \Cref{proof:idco}.
\qed
\end{lemma}
\noindent
\Cref{idco} shows that fixing $x_i$, the  payoff function $\pi_i(\cdot,x_i)$  is 
identified as  a constant, or identified up to location and scale, where the scale could be negative. In particular, if $\mathcal R_i(x_i)$ has the largest rank ${2^{I-1}}-1$, there are no strategic effects. Such a rank condition can be achieved when $X_{-i}$ only contains discrete variations. In particular, when $I=2$, the  $2^{I-1}-1$ rank of ${\mathcal R}_i(x_i)$ equals one, and can be achieved if $\pi_i(0,x_i)=\pi_i(1,x_i)$, and $X_j$ is a binary variable such that conditional on $X_i=x_i$, $X_j$ induces a single variation in player $j$'s (marginal) choice probability.

Under an additional assumption (i.e. assumption V below), we identify the existence of strategic effects. In addition, we can identify the sign of  $\pi_i(a_{-i},x_i)-\pi_i(a'_{-i},x_i)$. When players' private signals are independent,  \cite{de2012inference} develop a special approach for nonparametrically identifying the signs of the strategic effects by exploiting the identification power of multiple equilibria.  In contrast, our approach relies on assumptions E, ER while being applicable when there is only one (m.p.s.) equilibrium.

\begin{MV}[Variations in Marginal Choice Probabilities] \label{strategic}
Fix $x_i\in\mathscr S_{X_i}$. There exist $\alpha,\alpha'\in\mathscr S_{\alpha(X)|X_i=x_i}$ such that $0<\alpha_i\neq \alpha'_i<1$   and $(\alpha_i,\alpha'_{-i})\in\mathscr S_{\alpha(X)}$.
\end{MV}

\begin{proposition}
\label{idco_new}
Suppose $S\in\mathcal M_2$. Fix $x_i\in\mathscr S_{X_i}$.  Then $ \pi_i(\cdot,x_i)$ varies on $\mathcal A_{-i}$ if   $\mathscr S_{\alpha_i(X)|X_i=x_i}$ is not a singleton. Moreover, suppose assumptions SC, ER and V hold. If the rank of ${\mathcal R}_i(x_i)$ is $2^{I-1}-2$, then the sign of $ \pi_i(a_{-i},x_i)-\pi_i(a^0_{-i},x_i)$  is identified for each $a_{-i}\in\mathcal A_{-i}$.
\proof
See \Cref{proof:idco_new}.
\qed
\end{proposition}


\noindent
 When assumption V holds, the rank condition in \Cref{idco_new} requires  that $X_{-i}$ contain at least one continuous random variable such that, conditional on $X_i=x_i$ and $\alpha_i(X)=\alpha_i$, there are sufficient variations in $\sigma^*_{-i}(\cdot|X,u_i^*(X))$  by varying $X_{-i}$. Such a rank condition  is related to \cite{pesendorfer2003identification} and \cite{bajari2010estimating} under semiparametric settings. 
 
Alternatively, it is worth pointing out that we can identify the payoffs in model $\mathcal M_2$ up to location and scale by using the single--index structure suggested in \Cref{lemma1}. To see this,  let $\overline\pi_i(a_{-i},x_i)=\pi_i(a_{-i},x_i)-\pi_i(a^0_{-i},x_i)$. Then
\begin{multline}
\label{alter_iden_h}
 \mathbb{E}\left(Y_i|X_i=x_i, \Sigma^*_{-i}(X)=\Sigma^*_{-i}(x)\right)
 \\
 =F_{U_i}\Big(\pi_i\left(a_{-i}^0,x_i\right)+\sum_{a_{-i}\in\mathcal{A}/\{ a_{-i}^0\}} \overline\pi_i(a_{-i},x_i)\times \sigma^*_{-i}\big(a_{-i}|x,u^*_i(x)\big)\Big).
\end{multline} 
 Similarly to \cite{powell1989semiparametric}, we can identify $\overline\pi_i(\cdot,x_i)$ up to scale by
differentiating (\ref{alter_iden_h}) with respect to $\Sigma^*_{-i}(x)$.
Thus $\overline\pi_i(\cdot,x_i)$ is identified up
to scale.
This identification strategy involves an additional support condition on $\mathscr S_{\Sigma^*_{-i}(X)|X_i=x_i}$ for taking the derivative, i.e., conditional on $X_i$,  the random vector $\Sigma^*_{-i}(X)$ has a convex support with nonempty interior. See \citet[][Assumption 1]{powell1989semiparametric}

To identify the payoffs up to a single location and scale, we introduce a normalization.  Similar to \cite{matzkin2003nonparametric}, our normalization is imposed on the payoff functions at some $x^*_i\in\mathscr S_{X_i}$.

\begin{PN}[Payoff Normalization]
\label{normlization_payoff}
We set $\pi_i\left(a_{-i}^0,x^*_i\right)=0$ and $ \|\pi_i(\cdot,x^*_i)\|=1$ for  some $ x^*_i\in\mathscr S_{X_i}$ satisfying (i) assumption V and (ii) the rank of ${\mathcal R}_i(x^*_i)$ is $2^{I-1}-2$.\footnote{W.l.o.g., we set $a^0_{-i}=(0,\cdots,0)\in\mathcal A_{-i}$. Note that $ \|\pi_i(\cdot,x^*_i)-\pi_i(a^0_{-i},x^*_i)\|\neq 0$ because of the non--degeneracy of the support $\mathscr S_{\alpha_i(X)|X_i= x^*_i}$.} 
\end{PN}

Let $S\in\mathcal M_2$.  Suppose that assumptions SC, ER and N hold. By \Cref{idco}, the payoffs $\pi_i(\cdot,x^*_i)$ is point identified. By (\ref{eqcond}), $Q_{U_i}$ is identified on the support $\mathscr S_{\alpha_i(X)|X_i=x_i^*}\bigcap (0,1)$. Further, for each $x_i\in\mathscr S_{X_i}$,  suppose that the rank of ${\mathcal R}_i(x_i)$ equals $2^{I-1}-2$, and that $\mathscr S_{\alpha_i(X)|X_i=x_i}\bigcap\mathscr S_{\alpha_i(X)|X_i=x_i^*}\bigcap (0,1)$ contains two elements $\alpha_i,\alpha'_i\in(0,1)$. By  \Cref{idco}, $\pi_i(\cdot,x_i)$ is  identified up to location and scale. Note that the quantiles $Q_{U_i}(\alpha_i)$ and $Q_{U_i}(\alpha'_i)$ are known since $\alpha_i,\alpha'_i\in \mathscr S_{\alpha_i(X)|X_i=x_i^*}\bigcap (0,1)$. Therefore, we can determine the location and scale of $\pi_i(\cdot,x_i)$ from the following two equations
\begin{gather*}
\sum_{a_{-i}\in\mathcal A_{-i}} \pi_{i}(a_{-i},x_i)\times \mathbb E \left[\sigma^*_{-i}(a_{-i}|X,u^*_i(X))|X_i=x_i,\alpha_i(X)=\alpha_i\right]=Q_{U_i}(\alpha_i);\\
\sum_{a_{-i}\in\mathcal A_{-i}} \pi_{i}(a_{-i},x_i)\times \mathbb E \left[\sigma^*_{-i}(a_{-i}|X,u^*_i(X))|X_i=x_i,\alpha_i(X)=\alpha_i'\right]=Q_{U_i}(\alpha_i').
\end{gather*}
Moreover, we can identify $Q_{U_i}$ on the support $\mathscr S_{\alpha_i(X)|X_i\in\{ x^*_i, x_i\}}\bigcap (0,1)$. Repeating such an argument, we can show that $\pi_i(\cdot, x_i)$ can be point identified for all $x_i$s in a collection, denoted as $\mathbb C^\infty_i$, while $Q_{U_i}$ is identified on the support $\mathscr S_{\alpha_i(X)|X_i\in \mathbb C^\infty_i}\bigcap (0,1)$. 

\begin{definition}\label{def:collection}
Let the subset $\mathbb C^\infty_i$ in $\mathscr S_{X_i}$ be defined by the following iterative scheme.  Let $\mathbb C^0_i = \{x^*_i\}$. Then, for all $t\geq 0$, $\mathbb C^{t+1}_i$ consists of all elements $x_i\in \mathscr S_{X_i}$ such that at least one of the following conditions is satisfied: (i) $x_i\in \mathbb C^{t}_i$; (ii) ${\mathcal R}_i(x_i)$ has rank $2^{I-1}-2$ and there exists an $x'_i\in  \mathbb C^{t}_i$ such that $\mathscr S_{\alpha_i(X)|X_i=x_i}\bigcap\mathscr S_{\alpha_i(X)|X_i=x'_i}\bigcap (0,1)$ contains at least two different elements; and (iii) ${\mathcal R}_i(x_i)$ has rank $2^{I-1}-1$ and there exists an $x'_i\in  \mathbb C^{t}_i$ such that $\mathscr S_{\alpha_i(X)|X_i=x_i}\subseteq\mathscr S_{\alpha_i(X)|X_i=x'_i}\bigcap (0,1)$. 
\end{definition}

 In view of \Cref{idco},  condition (ii) in \Cref{def:collection} corresponds to the case where there are strategic effects. This case is the key to effectively expand the collection of $x_i$s in an iterative manner by enlarging $ \mathscr S_{\alpha_i(X)|X_i\in  \mathbb C^t_i}$ to $ \mathscr S_{\alpha_i(X)|X_i\in  \mathbb C^{t+1}_i}$. Note that  to exploit condition (ii), we implicitly assume that $X_{-i}$ contains at least one continuous random variable.

\begin{proposition}
\label{idpayoff}
Let $S\in\mathcal M_2$.  Suppose assumptions SC, ER and N hold. Then $\pi_i$ and $Q_{U_i}$ are point identified on the support $\mathcal A_{-i}\times\mathbb C^\infty_i$ and $\mathscr S_{\alpha_i(X)|X_i\in \mathbb C^\infty_i}\bigcap (0,1)$, respectively. 
\proof See \Cref{proof_idpayoff}
\qed
\end{proposition}

It is interesting to note that  our identification argument does not apply to a nonparametric single--agent binary response model, see e.g. \cite{matzkin1992nonparametric}.\footnote{In single--agent binary response models, \cite{matzkin1992nonparametric} establishes nonparametric identification results under  additional model restrictions (e.g., her assumptions W.2, W.4 and G.2).} This is because the support $\mathscr S_{\alpha_i(X)|X_i=x^*_i}$  is a singleton in a single--agent binary response model, i.e., we are always in the case of condition (iii) in  \Cref{def:collection}. In contrast, with interactions and exclusion restrictions, we can exploit  variations of $X_{-i}$ while controlling for $X_i$ to identify a set of quantiles of $F_{U_i}$.

Note that $\{\mathbb C^t_i:t\geq 1\}$ is an expanding sequence on the support of $\mathscr S_{X_i}$, which ensures that the limit $\mathbb C^\infty_i$ is well defined (and may not be bounded if $\mathscr S_{X_i}$ is unbounded).  The  domain and size of  $\mathbb C^\infty_i$ depend on the choice of $x^*_i$ as well as the variation of $\mathscr S_{\alpha_i(X)|X_i=x_i}$ across different  $x_i$s. Regarding the choice of the starting point $x^*_i$, intuitively we should choose it in a way such that $\mathbb C^\infty_i$ is the largest. However, it can be shown that for any $ x_i'$ satisfying  assumption N, if  $x_i'\in\mathbb C^\infty_i$, then we will end up with the same $\mathbb C^\infty_i$; otherwise $x'_i$ will lead to a non--overlapping set ${\mathbb C_i^\infty}'$.

The next corollary shows that the above iterative mechanism is not necessary if $x^*_i$ provides the largest variations in player $i$'s marginal choice probability conditional on $X_i$. 

\begin{PN}
(iii)  $\mathscr S_{\alpha_i(X)|X_i=x_i}\subseteq \mathscr S_{\alpha_i(X)|X_i=x_i^*}$ for all $x_i\in\mathscr S_{X_i}$. 
\end{PN}
\noindent
Assumption N-(iii) requires  $X_{-i}$ to have sufficient variations conditional on $X_i=x_i^*$, which is satisfied in various situations. For instance, this is the case when  $\mathscr S_{\alpha_i(X)|X_i=x^*_i}$ has full support $[0,1]$. See e.g. \cite{wan2010semiparametric,lewbel2012identification}.

\begin{corollary}
\label{corollary2}
Let $S\in\mathcal M_2$.  Suppose assumptions SC, ER and N (i) to (iii) hold. Then the results in \Cref{idpayoff} hold, where $\mathbb C^\infty_i=\{x_i\in\mathscr S_{X_i}:\ \text{Rank of }\ {\mathcal R}_i(x_i)\geq 2^{I-1}-2\}$.
\end{corollary}

There are normalizations other than assumption N.  For instance, we can normalize 
two quantiles of the marginal distributions. Specifically, for $\tau_{i1}, \tau_{i2}\in\mathscr S_{\alpha_i(X)|X_i=x^*_i}\bigcap(0,1)$,  we can  set the quantiles $Q_{U_i}(\tau_{i1})$ and $Q_{U_i}(\tau_{i2})$ at some values, as long as (strict) monotonicity is satisfied. \Cref{idpayoff} still holds. Second, we can use the usual mean/variance normalization in binary variable models. This is possible under a full support condition. Namely, suppose ${\mathcal R}_i(x^*_i)$ has rank $2^{I-1}-2$ and $(0,1)\subseteq\mathscr S_{\alpha_i(X)|X_i=x_i^*}$. Then we can set $\mathbb E (U_i)=0$ and $\text{Var} (U_i)=1$. Specifically, by \Cref{idco},  $\pi_i$ is identified up to location and scale. Hence, all the quantiles $Q_{U_i}$ are identified up to location and scale by (\ref{eqcond}). The latter are determined by the mean and variance normalization.  

Lastly, we note that given the identification of the joint distribution of types, we might be interested in some additional structures on the error terms. For instance, suppose the private signals are affiliated in the sense of \cite{milgrom1982theory} as when $U_i=\xi+\epsilon_i$ where $\xi$ is a common shock to all players and $\epsilon_i$ are iid (across players) idiosyncratic errors.  Following \cite{li1998nonparametric}, we can further deconvolute the joint distribution $F_U$ to identify the  marginal distributions of $\xi$ and $\epsilon_i$.

\subsection{Identification of $\mathcal M_3$}
Though not our focus of interest, $\mathcal M_3$ is nonparametrically identified in general. The argument does not essentially differ from that of $\mathcal M_2$: assumption I only relaxes the support condition for identification of $\sigma^*_{-i}$ in \Cref{iden0}. We illustrate this in the next lemma.  
\begin{lemma}
\label{iden_m4}
Let $S\in\mathcal M_3$. Fix $x\in\mathscr S_X$. Then $\sigma^*_{-i}(\cdot|x,u^*_i(x))$ is identified by 
\begin{equation*}
\sigma^*_{-i}(a_{-i}|x,u^*_i(x))
= \Pr \left(Y_{-i}=a_{-i}|X=x \right).
\end{equation*} 
\end{lemma}
The proof is straightforward, hence omitted.  By  \Cref{coro1}, we can also show that $\sigma^*_{-i}(a_{-i}|x,u^*_i(x)) = \Pr \left(Y_{-i}=a_{-i}|\alpha(X)=\alpha(x) \right)$. Similar results can be found in  \cite{pesendorfer2003identification,aguirregabiria2007sequential,bajari2010estimating}, among others. Further,  the identification of $\pi_i$ and $Q_{U_i}$ in $\mathcal M_3$ follows \Cref{idco} and \Cref{idpayoff} under assumptions ER and N. In particular, if $\mathscr S_{\alpha_i(X)|X_i=x_i}$ is not a singleton, then the rank condition requires that $X_{-i}$ contain at least one continuous random variable to ensure there are sufficient variations in $\sigma^*_{-i}(\cdot|X,u_i^*(X))$ conditional on $X_i$ and $\alpha_i(X)$ by varying $X_{-i}$. For a recent contribution using special regressors, see \cite{lewbel2012identification}. 

\section{Discussion}

In this section, we consider four issues related to our identification analysis. First, we illustrate how to nonparametrically estimate $\mathcal M_2$ based on the identification strategy established in Section 4.2. Second, without assumption ER, we examine the partial identification of $\mathcal M_2$. Third, we  relax the single equilibrium assumption, i.e., we allow for multiple m.p.s. BNEs in the DGP, under which the observed data is a mixture of distributions from all these equilibria. Fourth, we discuss the issue of unobserved heterogeneity. 

\subsection{A Sketch of Nonparametric Estimation}
To demonstrate how our nonparametric identification results can be used for estimation, we provide a sketch of a simple estimation procedure of a structure $\left[\pi; \{Q_{U_i}\}_{i=1}^I;C_{U}\right]$ in $\mathcal M_2$. A full development of  nonparametric inference is beyond the scope of this paper. 

For simplicity, we maintain the conditions in \Cref{corollary2}. Suppose the researcher observes an iid random sample $\{(X'_1,Y'_1)',\cdots,(X'_n,Y'_n)'\}$, where $X_t=(X'_{1t},\cdots,X'_{It})'$ and $Y_t=(Y_{1t},\cdots,Y_{It})'$ for $t=1,\cdots,n$. Note that the number of players $I$ is assumed to be constant for expositional simplicity. Here we suggest a flexible two--stage estimation procedure using sieve methods. 

\noindent 
\textbf{Step 1: Estimate the copula function $C_U$.} For this step, we begin by estimating the marginal choice probability function $\alpha_i(\cdot)$. There are several alternative nonparametric methods for such a purpose. Here we use sieve methods; see \cite{chen2007large}. Let $F$ be a continuous distribution function with an interval support. For instance, we can choose $F=\Phi$, the standard Normal distribution. We then estimate $\alpha_i(\cdot)$ by $\hat \alpha_i(\cdot)=F(\hat \gamma_i(\cdot))$, where 
\[
\hat \gamma_i=\argsup_{\gamma_i\in \Gamma_n} \ \frac{1}{n}\sum_{t =1}^n \left\{Y_{it}\log F(\gamma_i(X_{t}))+(1-Y_{it}) \log \big[1-F(\gamma_i(X_{t}))\big]
\right\},
\]in which $\Gamma_n$ is a H$\ddot{\text{o}}$lder class of real valued smooth  basis functions mapping $\mathscr S_{X}$ to $\mathbb R$.

The estimation of $C_U$ follows \cref{iden_copula}. Note that with $C_U$ and $\alpha(X)$, we can obtain the conditional choice probability of $Y=a$ for any $a\in\mathcal A$ given $X$. Let $\Pr(Y=a|X)=G_I(a; C_U,\alpha(X))$ where $G_I$ is a known function depending on $I$, $C_U$ and $\alpha(X)$. For example, suppose $I=2$. We can show that 
\begin{align*}
G_2((1,1); C_U,\alpha)&=C_U(\alpha),\\
G_2((0,1); C_U,\alpha)&=C_U(1,\alpha_{2})-C_U(\alpha),\\ 
G_2((1,0); C_U,\alpha)&=C_U(\alpha_{1},1)-C_U(\alpha),\\ 
G_2((0,0); C_U,\alpha)&=1-C_U(\alpha_1,1)-C_U(1,\alpha_2)+C_U(\alpha).
\end{align*}
Further, let $\mathcal C_n$ be a H$\ddot{\text{o}}$lder class of ``p--smooth'' real valued  basis functions mapping $[0,1]^I$ to $\mathbb R$ for some $p> 1$, which can approximate any square--integrable function arbitrarily well. Using the sieve MLE method, we then define our copula estimator by 
\[
\hat C_U=\argsup_{C_U\in \mathcal C_n} \ \frac{1}{n}\sum_{t =1}^n \sum_{a\in\mathcal A}\mathbb 1(Y_t=a)\log  G_I(a; C_U,\hat \alpha(X)).
\]
Consistency and asymptotic distribution of $\hat \alpha(\cdot)$ and $\hat C_U$ can be obtained from e.g.  \cite{chen2007large}.\footnote{Alternatively,  \eqref{iden_copula} can be used to propose a kernel regression estimator of the copula with generated regressors $\hat{\alpha}(X)$. See Mammen, Rothe and Schienle (2012).} As the function estimator $\hat C_{U}$ might not be a proper copula, we can modify $\hat C_U$ by using a rearrangement approach similar to \cite{chernozhukov2010rearranging}. Care must be taken given that $C_U$ is a multivariate CDF.

\hspace{6pt}

\noindent
\textbf{Step 2: Estimate the payoff functions $\pi_i$ and quantile functions $Q_{U_i}$.} Using \Cref{iden0}, we first estimate the equilibrium beliefs $\sigma^*_{-i}$ by
\[
\hat \sigma^*_{-i}(a_{-i}|X_t,u_i^*(X_t))=\frac{\partial G_I((1,a_{-i}); \hat C_U,\alpha)}{\partial \alpha_i}\big|_{\alpha=\hat \alpha(X_t)}.
\]

Next, we estimate $\pi_i$ and $Q_{U_i}$ from \cref{eqcond}. Suppose $Q_{U_i}\in \mathcal L^2(0,1)$. Then, for a given complete orthonormal sequence $\{\psi_k, k\geq 1\}$ in $\mathcal L^2(0,1)$, we have
\[
Q_{U_i}(\alpha)= \sum_{k=1}^\infty q^*_k \cdot \psi_k(\alpha), \ \ \forall \alpha\in \mathscr S_{\alpha_i(X)},
\]where $q^*_k=\int_0^1 \psi_k(s)\cdot Q_{U_i}(s) ds$. Let $\mathcal Q_n=\big\{\sum_{k=1}^{K_n}q_k\psi_k: q_k\in\mathbb Q_k\big\}$ be a sieve space depending on the sample size, where $\mathbb Q_k\subset \mathbb R$ is compact. Note that (\ref{eqcond}) implies that $\pi_i$ is identified up to $Q_{U_i}$, i.e.,
\[
 \pi_i(\cdot,x_i)=\left\{\mathbb E \left[{\Sigma}^*_{-i}(X){\Sigma_{-i}^{*'}}(X)\big|X_i=x_i\right]\right\}^{-1}\mathbb E \left[{\Sigma}^*_{-i}(X) Q_{U_i}(\alpha_i(X))\big|X_i=x_i\right].
\]
Hence, our estimator of $Q_{U_i}$ is defined as follows:
\begin{multline*}
\hat Q_{U_i}
=\arginf_{Q_{U_i}\in \mathcal Q_n} \sum_{t=1}^n\Big[\sum_{a_{-i}\in\mathcal A_{-i}} \tilde \pi_i(a_{-i},X_{it}|Q_{U_i})\cdot \hat \sigma^*_{-i}(a_{-i}|X_t,u^*_i(X_t))-Q_{U_i}(\hat \alpha_i(X_t))\Big]^2\\
s.t.\ \ \tilde \pi_i\left(a_{-i}^0,x^*_i|Q_{U_i}\right)=0,\ \text{ and } \ \ \|\tilde \pi_i(\cdot,x^*_i|Q_{U_i})\|=1,
\end{multline*} where $ \tilde \pi_i(\cdot,\cdot|Q_{U_i})$ is a functional of $Q_{U_i}=\sum_{k=1}^{K_n}q_k\psi_k$:
\begin{multline*}
\tilde \pi_i(\cdot,X_{it}|Q_{U_i})\\
\equiv\left[\sum_{s=1}^n{\hat\Sigma}^*_{-i}(X_s){{\hat\Sigma}_{-i}^{*'}}(X_s)K\Big(\frac{X_{is}-X_{it}}{h}\Big)\right]^{-1}\left[\sum_{s=1}^n{\hat\Sigma}^*_{-i}(X_s) Q_{U_i}(\hat \alpha_i(X_s))K\Big(\frac{X_{is}-X_{it}}{h}\Big)\right]\\
=\sum_{k=1}^{K_n}q_k\left\{\left[\sum_{s=1}^n{\hat\Sigma}^*_{-i}(X_s){{\hat\Sigma}_{-i}^{*'}}(X_s)K\Big(\frac{X_{is}-X_{it}}{h}\Big)\right]^{-1}\left[\sum_{s=1}^n{\hat\Sigma}^*_{-i}(X_s) \psi_k(\hat \alpha_i(X_s))K\Big(\frac{X_{is}-X_{it}}{h}\Big)\right]\right\},
\end{multline*} where $K$ and $h$ are a kernel function and bandwidth, respectively. Then, we let $\hat \pi_i(\cdot,x_i)=\tilde \pi_i(\cdot,x_i|\hat Q_{U_i})$. The proposed estimation procedure is easy to implement. Its precise asymptotic properties can be derived using the functional delta method in \cite{van1996weak}. 
As the quantile function estimator $\hat Q_{U_i}$ might not be strictly increasing, we can use \cite{chernozhukov2010quantile}'s rearrangement approach to modify it, or choose a shape preserving sieve as suggested by e.g. \cite{chen2007large}.

\subsection{Partial Identification}\label{sec:partial identification}
In this subsection, we study the partial identification of the game primitives when there are no exclusion restrictions, i.e. when assumption ER does not hold.
It is worth emphasizing that the lack of point identification of a structure here is not due to multiple equilibria, but to the lack of identifying restrictions, i.e.,
the exclusion restrictions and the rank conditions.   This is similar to, e.g., \cite{shaikh2011partial} who study partial identification of the average structural function in a triangular model without imposing a
restrictive support condition.   
When there are multiple m.p.s. BNEs, we still maintain the assumption of a single equilibrium being played for generating the distribution of observables. 

By the same argument as for the identification of  $\mathcal M_2$, the copula function $C_U$ is  point--identified on the extended support $\mathscr S^e_{\alpha(X)}$.  Let $\mathscr C$ be the set of strictly increasing (on $(0,1]^I$) and continuously differentiable copula functions mapping $[0,1]^I$ to $[0,1]$. Then, the identification region of $C_U$ can be characterized by 
\[
\mathscr C_I=\Big\{\widetilde C_U\in\mathscr C: \widetilde  C_{U}(\alpha)= C_{U}(\alpha), \ \ \forall \alpha\in\mathscr S^e_{\alpha(X)}\Big\}.
\]
For each $ \widetilde C_U\in\mathscr C_I$, suppose we set $\widetilde F_{U_i}$ to be the uniform distribution on $[0,1]$ and $\tilde\pi_i(\cdot, x)=\alpha_i(x)$. Clearly, the constructed structure $\big[\tilde \pi;  \widetilde F_U\big]$ is observationally equivalent to the underlying structure. Thus, $\mathscr C_I$ is the sharp identification region for $C_U$.

Next, we turn to the set identification of  the quantile function $Q_{U_i}$. By assumption R, $Q_{U_i}$ belongs to the set of strictly increasing and continuously differentiable functions mapping  $[0,1]$ to $\mathbb R$, denoted as $\mathscr Q$.  The next lemma shows that $\mathcal M_2$  imposes no restrictions on $Q_{U_i}$ and its identification region is $\mathscr Q$. 
\begin{lemma}
\label{lem_set_1}
Let $S\in\mathcal M_2$. For any $(\widetilde Q_{U_1},\cdots, \widetilde Q_{U_I})\in\mathscr Q^I$,  there exists an observationally equivalent structure $\widetilde S\in\mathcal M_2$  with the marginal quantile function profile $(\widetilde Q_{U_1},\cdots, \widetilde Q_{U_I})$. 
\proof  See \Cref{proof_lem_set_1} \qed
\end{lemma}

Now we discuss the sharp identification region for  $\pi_i$. Let $\mathscr G$ be the set of functions mapping $\mathcal A_{-{i}}\times \mathscr S_X$ to $\mathbb R$. 
\begin{proposition}
\label{set_id_1}
Let $S\in\mathcal M_2$. Suppose assumption SC holds. Then the sharp identification region is given by $\left\{\big[\widetilde \pi; \{\widetilde Q_{U_i}\}_{i=1}^I; \widetilde C_U\big]: (\widetilde Q_{U_i}, \widetilde C_U)\in(\mathscr Q, \mathscr C_I), \  \widetilde \pi\in\Theta_I\big(\{\widetilde Q_{U_i}\}_{i=1}^I, \widetilde C_U\big)\right\}$, where 
\begin{multline*}
\Theta_I(\{\widetilde Q_{U_i}\}_{i=1}^I, \widetilde C_U)\equiv\Big\{\widetilde \pi\in \mathscr G^I: (a) \text{ for all }x\in\mathscr S_X \text{ and }i, \\
 \widetilde Q_{U_i}(\alpha_i(x))=\sum_{a_{-i}\in\mathcal A_{-i}}\widetilde \pi_i(a_{-i},x)\times \sigma^*_{-i}(a_{-i}|x,u^*_i(x));  (b) \text{ for any m.p.s. profile } \delta: \\
{ \mathbb E}_\delta\left[\widetilde\pi_i\left(Y_{-i}, X\right)|X=x,U_i=\widetilde Q_{U_i}(\alpha_i)\right] - \widetilde Q_{U_i} (\alpha_i) \ \text{ is weakly monotone in }\alpha_i\in(0,1)
  \Big\}.
\end{multline*}
\proof See \Cref{proof_set_id_1}. \qed
\end{proposition}
\noindent
In the definition of $\Theta_I$, condition (a) requires that  $\pi_i(\cdot,x)$ should belong to a hyperplane, for which the slopes are given by the identified beliefs $ \Sigma^*_{-i}(x)$; condition (b) does not impose much restriction on the structural parameters. Clearly,   $\Theta_I$ is nonempty and convex.\footnote{ To see the nonemptiness, we can simply take  $\widetilde \pi_i(\cdot, x)=\widetilde Q_{U_i}(\alpha_i(x))$.}

The identification region is unbounded and quite large. To see this, fix an arbitrary non--negative function $\kappa_i(x)\geq 0$. Let $\psi_i:\mathbb R\rightarrow \mathbb R$ satisfy: (i) $\psi_i$ is a continuously differentiable and strictly increasing function; and (ii) for all $x$, $\kappa_i(x) Q_{U_i}(\alpha_i)-\psi_i(Q_{U_i}(\alpha_i))$ is weakly decreasing in $\alpha_i\in (0,1)$. Note that condition (ii) is equivalent to: $\inf_{u_i\in\mathscr S_{U_i}}\psi_i'(u_i)\geq \sup_{x\in\mathscr S_X}\kappa_i(x)$. Clearly, there are plenty of  choices for such a function $\psi_i$. Let further $\widetilde Q_{U_i}=\psi_i(Q_{U_i})$ and $\tilde \pi_i(a_{-i},x)= \xi_i(x)+ \kappa_i(x)\times  \pi_i(a_{-i}, x)$, in which $\xi_i(x)=\psi_i(Q_{U_i}(\alpha_i(x)))-\kappa_i(x)\times Q_{U_i}(\alpha_i(x))$.  Then, it can be verified that the constructed structure $[\widetilde \pi; \{\widetilde Q_{U_i}\}_{i=1}^I; \widetilde C_U\big]$ belongs to the identified set.\footnote{Note that the payoff normalization imposed in \Cref{idpayoff} is not helpful to bound the payoffs, since it applies only at one point $x^*_i\in\mathscr S_{X_i}$.  Specifically, if one imposes a similar normalization  on $\pi_i(\cdot, x^*)$ for some $x^*\in\mathscr S_X$, we would need to restrict the monotone mapping $\psi_i$ to satisfy $\psi_i(\alpha_i(x^*))=\alpha_i^*(x)$ when constructing an observational equivalence structure. Nevertheless, without  assumption ER,
the unboundedness of the partially identified set still holds for all $x\in S_X$ satisfying $\alpha_i(x)\neq \alpha_i(x^*)$.} To narrow down the identification region, additional restrictions need to be introduced. Instead of imposing assumption ER, an alternative approach is to make assumptions on the payoff functional form. For instance,   \cite{de2012inference} set $\pi_i(a_{-i}, x)=\pi^*_i(x) + g_i(a_{-i}) \times h^*_i(x) $, where $g_i$ is a function known to all players as well as to the econometrician, and $(\pi^*_i,h^*_i)$ are structural parameters in their model. Then,  \Cref{set_id_1}--(a) becomes:   for all $x\in\mathscr S_X $ and $i$,
\[
 \widetilde Q_{U_i}(\alpha_i(x))=\tilde \pi_i^*(x)+\tilde h_i^*(x)\times \sum_{a_{-i}\in\mathcal A_{-i}}g_i(a_{-i})\times \sigma^*_{-i}(a_{-i}|x,u^*_i(x)),
 \]which imposes a linear restriction on $\tilde \pi_i^*(x)$ and $\tilde h_i^*(x)$ by noting that $ \sum_{a_{-i}\in\mathcal A_{-i}}g_i(a_{-i})\times \sigma^*_{-i}(a_{-i}|x,u^*_i(x))$ is identified under the conditions in \Cref{iden0}. Moreover,   \Cref{set_id_1}--(b) imposes an additional restriction on the copula function $\widetilde C_U$.

When $\mathscr S_{\alpha(X)}=(0,1)^I$, $\mathscr C_I$ degenerates to the singleton $\{C_U\}$. In this case, the sharp identification region for $(\pi,\{Q_{U_i}\}_{i=1}^I)$ can be characterized in a more straightforward manner:
\begin{multline*}
\Theta_I^*=\Big\{(\widetilde \pi,\{\widetilde Q_{U_i}\}_{i=1}^I )\in  \mathscr G^I\times \mathscr Q^I: (a')  \text{ for all }x\in\mathscr S_X \text{ and } i, \\
\widetilde Q_{U_i}(\alpha_i(x))=\sum_{a_{-i}\in\mathcal A_{-i}}\widetilde \pi_i(a_{-i},x)\times \sigma^*_{-i}(a_{-i}|x,u^*_i(x));
(b') \text{ and for all } \alpha_{-i}\in[0,1]^{I-1},   \\
\sum_{a_{-i}\in\mathcal A_{-i}}\widetilde \pi_i(a_{-i},x)\times   \sigma^\alpha_{-i}(a_{-i}, \alpha_{-i},\alpha_i)- \widetilde Q_{U_i} (\alpha_i) \ \text{ is weakly monotone in }\alpha_i\in(0,1)
  \Big\},
\end{multline*}
where $ \sigma^\alpha_{-i}(a_{-i},\alpha_{-i},\alpha_i)=\Pr_{C_U}(C_j\leq \alpha_j\ \forall  a_j=1; C_j> \alpha_j\ \forall  a_j=0 |C_i=\alpha_i)$.

\subsection{Multiple Equilibria in DGP}
The problems raised by multiple equilibria have a long history in economics. See e.g. \cite{jovanovic1989observable} for empirical implications of multiple equilibria, and  \cite{morris2001rethinking} for a recent discussion in macroeconomic modeling. The static games literature have struggled with difficulties arising from equilibrium multiplicity since the mid 1980s. See e.g. \cite{bjorn1984simultaneous}. Researchers have developed essentially three approaches. In the first approach, one assumes there is a single equilibrium in the DGP as we do. See e.g. \cite{aguirregabiria2013recent} for a survey.  Sometimes this assumption is satisfied when the model admits a unique equilibrium.  \cite{pesendorfer2003identification} provide empirical justifications for this assumption, in particular, when data come from the same game repeatedly played across different time periods. See also e.g. \cite{bajari2010estimating}. A more sophisticated solution is to identify a subset in the support of covariates that admit a unique equilibrium. See \cite{xu2010estimation} in a parametric setting.

In the second approach, a seminal paper by \cite{tamer2003incomplete} introduces partial identification analysis in a discrete game of complete information. This allows us to bound the parameters of interest without specifying which equilibrium is chosen. In a parametric model, \cite{aradillas2008identification} obtain inequality constrains by exploiting level--k rationality in either a complete or incomplete information framework. In a semiparametric setting with incomplete information, \cite{wan2010semiparametric} develop upper/lower bounds for equilibrium beliefs to achieve point identification of payoff parameters under a full support condition on regressors. In the third approach, one introduce a probability distribution $\lambda$ over the set of  equilibria. See e.g.  \cite{bjorn1984simultaneous,bajari2010identification} for complete information games, and  \cite{aguirregabiria2007sequential} for incomplete information games in parametric settings.  


In general,  the issue of multiple equilibria is a largely unexplored area of research in a nonparametric framework, which is considered in this paper.  We first address how to detect multiple equilibria in our setting. Next we discuss the problem of identification in the presence of multiple equilibria.  Our discussion below focuses on $\mathcal M_2$, since $\mathcal M_1$ imposes almost no restrictions by  \Cref{lemma_oes_1}.\footnote{In $\mathcal M_3$ or $\mathcal M_3'$, multiple equilibria can be detected by testing the conditional independence as shown in \Cref{ration_M_4,corolary1}. See also \cite{de2012inference}.}  

In $\mathcal M_2$, we can detect multiple equilibria from the model restrictions derived in \Cref{coro1}, specifically, restrictions R1 and R2. This is because, in the presence of multiple equilibria in the data, R1 and/or R2 are violated in general. Moreover, if  assumption ER holds, \eqref{eqcond} introduces additional model restrictions, providing stronger  power to detect the existence of multiple equilibria. To see this,  we first fix $X_i=x_i$. Under assumption ER, \eqref{eqcond} can be rewritten as
\begin{equation}
\label{eqcond_er}
\sum_{a_{-i}\in\mathcal A_{-i}}\pi_i(a_{-i},x_i)\times \sigma^*_{-i}(a_{-i}|x,u^*_i(x))-Q_{U_i}(\alpha_i(x))=0,
\end{equation}for all $x\in\mathscr S_{X|X_i=x_i}$. Suppose $\mathscr S_{\alpha_i(X)|X_i=x_i}$ is not a singleton. Then, there are strategic effects.  Conditional on $\alpha_i(X)=\alpha_i\in\mathscr S_{\alpha_i(X)|X_i=x_i}$ further, the random vector $ \Sigma^*_{-i}(X)$, which is a probability mass function on $\mathcal A_{-i}$, has to be distributed on  a hyperplane in $\mathbb R^{2^{I-1}}$. In addition, the slope $\pi_i(\cdot,x_i)$ of the hyperplane remains constant as $\alpha_i$ varies, while its intercept $Q_{U_i}(\alpha_i(x))$ strictly increases in $\alpha_i$. Because the equilibrium beliefs $\sigma^*_{-i}(\cdot|x,u^*_i(x))$ are identified under assumption SC, violations of these restrictions indicate the presence of multiple equilibria.   In the special case of $I=2$,  these restrictions imply that, conditional on $X_i=x_i$,  $\alpha_j(x)$ is a  monotone function of $\alpha_i(x)$. Violations of such a monotonicity indicates multiple equilibria. 

%
%

Next, to relax the single equilibrium assumption for identification analysis, we follow \cite{henry2010identifying} by introducing an instrumental variable $Z$, which does not affect players' payoffs, the distribution of types, or the set of equilibria in the game, but can effectively change the equilibrium selection. For each $x\in\mathscr S_X$, let $\mathscr E(x)$ be the set of m.p.s. BNEs in the game with $X=x$.  Note that $\mathscr E(x)$ could be an infinite collection and the number of equilibria depends on the value of $x$. To simplify, we assume that players only focus on a subset $\Gamma(x)$ of $\mathscr E(x)$ for the DGP, i.e., the set of equilibria that will be played in the data. We further assume that the number of elements in $\Gamma(x)$ is finite and bounded above by a constant $J$ ($J\geq 2$) for all $x$. 


Let $\lambda$ be a probability distribution $\{p^\lambda_1,\cdots,p^\lambda_J\}$ on the support $\{1,\cdots,J\}$ such that the $j$--th equilibrium occurs with probability $p^\lambda_j$. The distribution $\lambda$ may have some zero mass points, which means that the number of equilibria in $\Gamma(x)$ is strictly less than $J$. Essentially, $\lambda$ summarizes the mixture of equilibrium distributions arising from the equilibrium selection mechanism. Following  \cite{henry2010identifying}, we assume that the probability distribution $\lambda$ varies with $X$ and $Z$, where $Z$ is a vector of  instrumental variables that does not affect either $\mathscr E(X)$ or $\Gamma(X)$, but has influence on the equilibrium selection through $\lambda$. 

In \cite{henry2010identifying}, it is shown that the set of component distributions is partially identified in the space of probability distributions. For example, for $J=2$, the observed distribution $F_{Y|X=x}$ is  a convex combination of the two component distributions generated from the two equilibria in $\Gamma(x)$. Then, variations of the instrumental variable $Z$ cause the mixture distribution to move along a straight line in the function space of probability distributions. 
Further, we can point identify the set of distributions corresponding to $\Gamma(x)$ if $Z$ has sufficient variations. To see this, w.l.o.g. let $J=2$. For each $x\in\mathscr S_X$, suppose there exist some (unknown) $z,z'\in\mathscr S_{Z|X=x}$ such that $\lambda=(0,1)$ when $(X,Z)=(x,z)$ and $\lambda=(1,0)$ when $(X,Z)=(x,z')$. In the space of probability distributions,  the two equilibrium distributions  can be identified as the two extreme points of the convex hull (which is a straight line) of the collection of distributions $F_{Y|X=x, Z=z}$ for all $z\in\mathscr S_{Z|X=x}$.  Either one of them represents a probability distribution from a single m.p.s. BNE   given $x$, which thereafter provides the identification of the underlying game structure as we discussed in the identification section.

\subsection{Correlated Types vs Unobserved Heterogeneity with Independent Types}
Within a paradigm  where private signals are independent unconditionally or conditionally given $X$, a known approach for generating correlation among actions given $X$ is to introduce unobserved heterogeneity. In a fully parametric setting, \cite{aguirregabiria2007sequential} and \cite{grieco2011discrete}  introduce unobserved heterogeneity through some payoff relevant variables $\zeta$ publicly observed by all players, but not by the researcher. An important question is whether one can  distinguish this model from our model with correlated types. Because $\mathcal M_1$  can rationalize any distributions generated by a model with unobserved heterogeneity and independent types, we consider $\mathcal M_2$ below.

Consider the following payoffs with unobserved heterogeneity:
\[
\Pi_i(Y,X,\zeta, U_i)=\left\{\begin{array}{cc}\pi_i(Y_{-i},X, \zeta)-U_i,&\text{ if } Y_i=1,\\0,&\text{ if } Y_i=0,\end{array}\right.
\]where $\zeta$ is  a discrete variable that is unobserved to the researcher. Let the support of $\zeta$ be $\{z_1,\cdots, z_J\}$ and $p_j(x)=\Pr(\zeta=z_j|X=x)$. To simplify, we assume there are only two players with $U_1\bot U_2$ and $(U_1,U_2)\bot (X,\zeta)$. We assume further that there is only a single equilibrium in the DGP for every $(x,z_j)\in\mathscr S_{X\zeta}$, which is also assumed in our model $\mathcal M_2$. Then, the joint choice probability for the model with unobserved heterogeneity and independent types is 
\[
\mathbb E (Y_1Y_2|X=x)=\sum_{j=1}^J \mathbb E (Y_1Y_2|X=x,\zeta=z_j)\cdot p_j(x)\\
=\sum_{j=1}^J \alpha_1(x,z_j)\cdot \alpha_2(x,z_j)\cdot p_j(x) 
\]where $\alpha_i(x,z_j)=\mathbb E (Y_i|X=x,\zeta=z_j)$. Moreover, the marginal choice probability is
\[
\alpha_i(x)=\sum_{j=1}^J \alpha_i(x,z_j)\cdot p_j(x), \ \ \text{for } i=1,2,
\]where $\alpha_i(x)=\mathbb E (Y_i|X=x)$. 

We now argue that these joint and marginal choice probabilities can violate restriction R1 in \Cref{coro1}. Specifically, R1 requires that $\mathbb E (Y_1Y_2|X=x)$ be a function of $(\alpha_1(x),\alpha_2(x))$ only. The model with unobserved heterogeneity and independent types does not exclude the possibility that there exist $x,x'\in\mathscr S_X$ such that $\alpha_i(x)=\alpha_i(x')$ for $i=1,2$, but $\mathbb E (Y_1Y_2|X=x)\neq \mathbb E(Y_1Y_2|X=x')$. For instance, suppose $J=2$, $p_j(x)=p_j(x')=0.5$ for $j=1,2$, $\alpha_1(x,z_1)=\alpha_2(x,z_1)=0.4$, $\alpha_1(x,z_2)=\alpha_2(x,z_2)=0.6$, $\alpha_1(x',z_1)=0.3$, $\alpha_2(x',z_1)=0.7$, $\alpha_1(x',z_2)=0.7$, $\alpha_2(x',z_2)=0.3$. Therefore, $\alpha_1(x)=\alpha_1(x')=0.5$ and $\alpha_2(x)=\alpha_2(x')=0.5$, but $\mathbb E (Y_1Y_2|X=x)=0.26\neq \mathbb E (Y_1Y_2|X=x')=0.21$. Consequently, the model with unobserved heterogeneity and independent types can be distinguished from model $\mathcal M_2$. We have focused on R1 above, but monotonicity in $\alpha_i(X)$ (see R2) could be violated as well for similar reasons.

\section{Conclusion}
This paper studies the rationalization and identification 
of discrete games with correlated types within a fully nonparametric framework. Allowing for correlation across types is important in global games and in models with social interactions as it represents correlated information and homophily, respectively. Regarding rationalization, we show that our baseline game--theoretical model $\mathcal M_1$ with a single m.p.s. BNE in the DGP  does not impose any essential restrictions on observables, and hence is not testable in view
of players' choice probabilities only.  We also show that exogeneity is testable, because R1--R3 in \Cref{coro1} characterize all the restrictions imposed by exogeneity. For instance, we can view R1 as a regression of the joint choices on covariates  that depends on the latter only through the marginal choice probabilities. Thus, to test R1 we can extend \cite{fan1996consistent} and \cite{lavergne2000nonparametric} significance tests by allowing for  estimation of the marginal choice probabilities. Moreover, R2 is a monotonicity restriction that can be  tested by testing the  convexity of its integral, see e.g., \cite{delgado2012distribution}.\footnote{In the context of finite normal form games, the Quantal Response Equilibrium has an identical structure to BNE in our setting.  In a semiparametric setting,  \cite{melo2014testing} obtain model restrictions characterized by monotonicity  that are similar to R1, and then propose a moment inequality test.}

Model   $\mathcal M_3$ is mostly adopted in empirical work within a parametric or semiparametric setting. We show that all its restrictions reduce to the mutual independence of choices conditional on covariates. This can be tested by using conditional
independence tests, see e.g. \cite{su2007consistent,su2008nonparametric}.  Moreover,  \Cref{ration_M_4} and \Cref{corolary1} show that the same restriction characterizes model $\mathcal M_3'$ which only assumes mutual independence of types conditional on covariates and a single (not necessary monotone) BNE in the DGP. These two assumptions seem to be unrelated, but actually are  two sides of the same coin. Maintaining a single equilibrium in the DGP, we can use the mutual independence of choices given covariates to test  mutual independence of types which is widely assumed in the literature. On the other hand, maintaining mutual independence of types, we can use the same mutual independence of choices to test for a single equilibrium versus multiple equilibria.  See, e.g., \cite{de2012inference}.

It is worth noting that the above tests do not rely on identification and consequently on the assumptions used to identify the primitives of the various models. In particular, we show that model $\mathcal M_2$ is identified up to a single location--scale normalization under  exclusion restrictions, rank conditions and a non--degenerate support condition. 
The  exclusion restrictions take the form of excluding part of a player's
payoff shifters from all other players' payoffs as frequently assumed
in the  literature. Specifically, the dependence of players' joint choices on the marginal choice probabilities identifies the dependence across types, while the dependence of a player's marginal choice probabilities on her equilibrium beliefs identifies her payoffs. Without exclusion restrictions, we show that the sharp identification region of players' payoffs is unbounded. 

Our identification results are useful for estimation of global games and social interaction models.  In a semiparametric setup, \cite{liu2012semiparametric} propose an estimation procedure for our model
$\mathcal M_2$ with linear payoffs, and establish the root--n consistency of the linear payoff coefficients.
A fully nonparametric estimation deserves to be studied following the estimation sketch given in Section 5.1.
Specifically, we could rely on the identification
results and propose sample--analog  estimators for the
players' payoffs and the joint distribution of private information. The equilibrium condition (\ref{eqcond}) is the key estimating equation. It has the nice feature to be partially linear, namely, linear in the payoffs and nonparametric in the quantile.  A
difficulty is to take into account the estimation of the beliefs of the player at the margin and the marginal choice probabilities. Part of the problem
could be addressed by using the recent literature on
nonparametric regression with generated covariates, see e.g. \cite{mammen2012nonparametric}. An important question is to determine the optimal (best) rate at which the primitives of $\mathcal M_2$ can be estimated from players' choices. \Cref{coro1} which characterizes all the restrictions imposed by $\mathcal M_2$ will be useful, see e.g. \cite{guerre2000optimal}.

\bibliographystyle{Chicago}
\bibliography{bibles}

\clearpage
\appendix
\section{Existence of m.p.s. BNEs}
\small

\subsection{Proof of \Cref{lemma1}}\label{proof_lemma1}

First, we show the existence of m.p.s. BNE. Assumptions G1--G6 of \citet{reny2011existence} are satisfied in our discrete game under assumption R. Moreover, by assumption M, when other players employ m.p.s., player $i$'s best response is  also a joint--closed set of m.p.s.. By  \citet[Theorem 4.1]{reny2011existence}, the conclusion follows.

We now show the second half. Fix $X=x$. Because $\sigma^*_{-i}(a_{-i}|x,u_i)$  are continuous in $u_i$ under assumption R,  then  $  \mathbb \sum_{a_{-i}} \pi_i(a_{-i},x)\sigma^*_{-i}(a_{-i}|x,u_i)-u_i$ is a continuously decreasing function in $u_i$. 

Suppose $\underline u_i(x)<u^*_i(x)<\overline u_i(x)$. It follows that
\[
 \mathbb \sum_{a_{-i}} \pi_i(a_{-i},x)\sigma^*_{-i}(a_{-i}|x,u_i^*(x))- u^*_i(x)=0.
\] Hence, conditional on $\underline u_i(X)<u^*_i(X)<\overline u_i(X)$, we have
\[
Y_i=\textbf 1\left[U_i\leq u^*_i(X)\right]=\textbf 1\left[U_i\leq \sum_{a_{-i}} \pi_i(a_{-i},X)\sigma^*_{-i}\big(a_{-i}|X,u^*_i(X)\big)\right].
\]  

Suppose $u^*_i(x)=\overline u_i(x)$. Then $ \mathbb \sum_{a_{-i}} \pi_i(a_{-i},x)\sigma^*_{-i}(a_{-i}|x,\overline u_i(x))- \overline u_i(x)\geq 0$, which implies that conditional on $u^*_i(X)=\overline u_i(X)$, there is
\[
Y_i=\textbf 1\left[U_i\leq  \overline u_i(X)\right]\leq \textbf 1\left[U_i\leq \sum_{a_{-i}} \pi_i(a_{-i},X)\sigma^*_{-i}\big(a_{-i}|X,\overline u_i(X)\big)\right].
\]
Because $\textbf 1\left[U_i\leq \overline u_i(X)\right]=1$ a.s., thus 
\[
Y_i=\textbf 1\left[U_i\leq \overline u_i(X)\right]=\textbf 1\left[U_i\leq \sum_{a_{-i}} \pi_i(a_{-i},X)\sigma^*_{-i}\big(a_{-i}|X,\overline u_i(X)\big)\right]\ \ \ a.s.
\] Similar arguments hold for the case $u^*_i(X)=\underline u_i(X)$.

\qed

\subsection{Existence of m.p.s. BNEs under primitive conditions}\label{existence_msbe}
\begin{definition}\label{upper}
a set $A\subseteq \mathbb{R}^d$ is upper if and only if its indicator function is non--decreasing, i.e., for any $x,y\in\mathbb{R}^d$, $x\in A$ and $x\leq y$ imply $y\in A$, where $x\leq y$ means $x_i\leq y_i$ for $i=1,\cdots,d$.
\end{definition}

\begin{PR}[Positive Regression Dependence]\label{pca}
For any $x\in\mathscr S_X$ and any upper set $A\subseteq\mathbb{R}^{I-1}$, the conditional probability $\Pr\left(U_{-i}\in A|X=x,U_i=u_i\right)$ is non--decreasing in $u_i\in\mathscr S_{U_i|X=x}$.
\end{PR}

\begin{SCP}[Strategic Complement Payoffs]\label{sca}
For any $x \in \mathscr S_X$ and $u_i\in\mathscr{S}_{U_i|X=x}$, suppose $a_{-i}\leq a'_{-i}$, then $\pi_i(a_{-i},x)\leq \pi_i(a'_{-i},x)$.
\end{SCP}

\begin{lemma}\label{msbe1}
Suppose  assumptions R, PRD and SCP hold.  For any $x\in\mathscr S_X$, there exists an m.p.s. BNE.  
\proof 
By \Cref{lemma1}, it suffices to show that assumption M holds.  Fix $x\in\mathscr S_X$. Given an arbitrary m.p.s. profile: for $i=1,\cdots,I$, $\delta_i(x,u_i)=\textbf{1}[u_i\leq u_i(x)]$, where $u_{i}(\cdot)$ is arbitrarily given. By assumptions PR and SCP, and \cite{lehmann1955ordered}, for any $u_i<u_i'$ in the support, we have 
\[
\mathbb E_\delta \left[\pi_i(Y_{-i}, X)|X=x, U_i=u_i'\right]\leq \mathbb E_\delta \left[\pi_i(Y_{-i}, X)|X=x, U_i=u_i\right].
\]
Thus, $\mathbb{E}_\delta\left[\pi_i(Y_{-i},X)|X=x,U_i=u_i\right]-u_i$ is a weakly decreasing function of $u_i$. 
\qed
\end{lemma}

\section{Rationalization}
\label{sectionAB}
\subsection{Proof of \Cref{lemma_oes_1}}\label{proof_lemma_oes_1}
Prove the ``only if part'' first: Proofs by contradiction. Let $F_{Y|X}$ be rationalized by $\mathcal M_1$, i.e., some $S\in\mathcal M_1$ can generate $F_{Y|X}$.  Fix $X=x$ and let equilibrium be characterized by $(u^*_1(x),\cdots, u^*_I(x))$. For some $a\in\mathcal A$, w.l.o.g., $a=(1,\cdots,1)$, suppose $\Pr(Y=a|X=x)=0$ and $\Pr(Y_i=a_i|X=x)>0$ for all $i$. It follows that $\Pr(U_1\leq u^*_1(x),\cdots,U_I\leq u^*_I(x)|X=x)=0$ and $\Pr(U_i\leq u_i^*(x)|X=x)>0$ for all $i$, which violates assumption R. Then $S\not\in\mathcal M_1$. Contradiction.

Proofs for the ``if part'': Fix an arbitrary $x\in\mathscr S_X$. First, we assume $\Pr(Y=a|X=x)>0$ for all $a\in\mathcal A$, which will be relaxed later. Now we construct a structure in $\mathcal M_1$ that will lead to $F_{Y|X}(\cdot|x)$. 

Let $\pi_i(a_{-i},x)=\alpha_i(x)$ for $i=1,\cdots,I$. Note that there is no strategic effect by construction and assumption M is satisfied.  Now we construct $F_{U|X}(\cdot|x)$. Let $ F_{U_i|X}(\cdot|x)$ be uniformly distributed on $[0,1]$. So it suffices to construct the copula function $C_{U|X}(\cdot|x)$ on $[0,1]^I$. We first construct $C_{U|X}(\cdot|x)$ on a finite sub--support: $\{\mathbb E (Y_1|X=x), 1\}\times \cdots\times\{\mathbb E (Y_I|X=x), 1\}$. Then we extend it to a proper copula function with the full support $[0,1]^I$.  Let $ C_{U|X}(\alpha_1,\cdots,\alpha_I|x)=\mathbb E (\prod_{j=1}^pY_{i_j}|X=x)$ where $i_1,\cdots,i_p$ are all the indexes such that $\alpha_{i_j}=\mathbb E (Y_{i_j}|X=x)$; while other indexes have $\alpha_k=1$. Because  $\Pr(Y=a|X=x)>0$  for all $a\in\mathcal A$, $ C_{U|X}(\cdot|x)$ is strictly increasing in each index  on the finite sub--support. Thus it is straightforward that we can extend $C_{U|X}(\cdot|x)$ to the whole support $[0,1]^I$ as a strictly increasing (on the support $(0,1]^I$) and smooth copula function. By construction, it is straightforward that the constructed structure can generate  $F_{Y|X}(\cdot|x)$.

When $\Pr(Y=a|X=x)=0$ for some $a$'s in $\mathcal A$.  By the condition in  \Cref{lemma_oes_1}, the conditional distribution of $Y$ given $X=x$ is degenerated in some indexes. W.l.o.g., let $\{1,\cdots,k\}$ be set of indexes such that $\Pr(Y_i=1|X=x)=0$ or $1$, and let $\{k+1,\cdots,I\}$ satisfy $0<\Pr(Y_i=1|X=x)<1$. Then let again $\pi_i(a_{-i},x)=\alpha_i(x)$ for $i=1,\cdots,I$. For player $i=k+1,\cdots,I$,  we can construct  a sub--copula function $C_{U_{k+1},\cdots,U_I|X}(\cdot|x)$ as described above such that $C_{U_{k+1},\cdots,U_I|X}(\cdot|x)$ is strictly increasing and smooth. Further, we can extend $C_{U_{k+1},\cdots,U_I|X}(\cdot|x)$ to a proper copula function having the full support $[0,1]^I$. Similarly, the constructed structure generates $F_{Y|X}(\cdot|x)$.
\qed

\subsection{Rationalizing All Probability Distributions}\label{rationalizing_all}

Suppose we replace assumption R with the following conditions in \cite{reny2011existence}:  For every $x\in\mathscr S_X$,

G.2. The distribution $F_{U_i|X}(\cdot|x)$ on $\mathscr S_{U_i|X=x}$ is atomless.

G.3. There is a countable subset $\mathscr S^0_{U_i|X=x}$ of $\mathscr S_{U_i|X=x}$ such that every set in $\mathscr S_{U_i|X=x}$  assigned
positive probability by $F_{U_i|X}(\cdot|x)$ contains two points between which lies a point in $\mathscr S^0_{U_i|X=x}$. 

Note that it is straightforward that assumptions G.1 and G.4 through G.6 in \cite{reny2011existence} are all satisfied in our discrete game because
the action space $\mathcal A$ is finite and the conditional distribution of
$U$ given $X=x$ has
a hypercube support in $\mathbb R ^I$. Thus, the conclusion in Lemma 1 still holds (i.e., existence of an m.p.s. BNE) under assumptions G.2, G.3 and M. Moreover, let $
\mathcal M'_1\equiv\left\{S:  \text{G.2, G.3 and M hold and a single m.p.s. BNE is played}\right\}$.
Then, we generalize \Cref{lemma_oes_1}.
\begin{lemma}
\label{lemma_oes_2}
Any conditional distribution $F_{Y|X}$ can be rationalized by $\mathcal M'_1$.
\proof
We prove by construction. Fix $x$. Let $\pi_i(a_{-i},x)=\alpha_i(x)$ for all $i$. Note that there is no strategic effect by construction and assumption M is satisfied.  Now we construct $F_{U|X}(\cdot|x)$. Let $[0,1]^I$ be the support of the distribution and partition it into $2^I$
disjoint events: $\bigotimes_{i=1}^I\{[0,\alpha_i(x)),[\alpha_i(x),1]\}$ 
\footnote{To have meaningful partition, it is understood that 
$\{[0,\alpha_i(x)),[\alpha_i(x),1]\}$ becomes 
$\{\{0\},(0,1]\}$ when $\alpha_i(x)=0$.}. Further, we define a conditional distribution $F_{U|X=x,U\in B_j}$ as a uniform distribution on $B_j$, where $B_j$ is the $j$--th event in the partition of the support. Moreover, let $\Pr(U\in B_j|X=x)=\Pr(Y=a(j)|X=x)$ where $a(j)\in\mathcal A$ and satisfies $a_i(j)=0$ if the $i$--th argument of event $B_j$ is $[\alpha_i(x),1]$, and $a_i(j)=1$ if the $i$--th argument is $[0,\alpha_i(x))$. With such construction, the marginal distribution of $U_i$ given $X=x$ is a uniform distribution on $[0,1]$ which satisfies assumptions G.2 
and G.3. It can be verified that the constructed structure  leads to $F_{Y|X}(\cdot|x)$.  
\qed
\end{lemma}

\subsection{Proof of \Cref{coro1}}
\label{proof_coro1}
\proof

We first show the "only if part''. Suppose that the distribution $F_{Y|X}(\cdot|\cdot)$ rationalized by $\mathcal M_1$  is derived from $\widetilde S=[\tilde{\pi};\tilde{F}_{U|X}]\in
\mathcal{M}_2$ . Then 
\begin{multline*}
\mathbb E\big(\prod_{j=1}^p Y_{i_j}|X\big)=\Pr\big(Y_{i_1}=1,\cdots, Y_{i_p}=1|X\big)\\
=\Pr\big(U_{i_1}\leq \tilde u^*_{i_1}(X), \cdots, U_{i_p}\leq \tilde u^*_{i_p}(X)|X\big)=\widetilde C_{U_{i_1},\cdots, U_{i_p}}\big(\alpha_{i_1}(X),\cdots, \alpha_{i_p}(X)\big).
\end{multline*}
Similarly, 
\[
\mathbb E\big(\prod_{j=1}^p Y_{i_j}|\alpha_{i_1}(X),\cdots, \alpha_{i_p}(X)\big)=\widetilde C_{U_{i_1},\cdots, U_{i_p}}\big(\alpha_{i_1}(X),\cdots, \alpha_{i_p}(X)\big).
\]Thus, we have condition R1.  Further, R2 and R3 obtain by the properties of the copula function $\widetilde C_{U_{i_1},\cdots,U_{i_p}}$. 

Proofs for the ``if part''.  For any $x\in\mathscr S_{X}$, let $\tilde\pi_i(\cdot,x)=\alpha_i(x)$.  Let $\tilde F_{U_i}$ denote the CDF of uniform distribution on $[0,1]$. For all $1\leq i_1<\cdots<i_p\leq I$, $(\alpha_{i_1},\cdots, \alpha_{i_p})\in \mathscr S_{\alpha_{i_1}(X),\cdots,\alpha_{i_p}(X)}$ and $x\in\mathscr S_X$,  define $\widetilde F_{U_{i_1},\cdots,U_{i_p}}(\cdot,\cdots,\cdot)$ as follows: for each $\alpha_{i_1},\cdots, \alpha_{i_p}\in\mathscr S_{\alpha_{i_1}(X),\cdots,\alpha_{i_p}(X)} $,
\[
\widetilde F_{U_{i_1},\cdots,U_{i_p}}(\alpha_{i_1}, \cdots,\alpha_{i_p})
=\mathbb E \Big[\prod_{j=1}^p Y_{i_j}\big|\alpha_{i_1}(X)=\alpha_{i_1},\cdots,\alpha_{i_p}(X)=\alpha_{i_p}\Big].
\]
Thus, we define $\widetilde F_U$ on the support $\{\alpha: (\alpha_{i_1}, \cdots,  \alpha_{i_p})\in\mathscr S_{\alpha_{i_1}(X),\cdots, \alpha_{i_p}(X)}; \text{other }\alpha_{i_j}=1\}$.

By  \Cref{lemma_oes_1}, we have that $
\widetilde F_{U_{i_1},\cdots,U_{i_p}, U_k}(\alpha_{i_1}, \cdots,\alpha_{i_p}, \alpha_k)<\widetilde F_{U_{i_1},\cdots,U_{i_p}}(\alpha_{i_1}, \cdots,\alpha_{i_p})$ for any $k\neq i_j$, $j=1,\cdots,p$, $\alpha_{i_j}>0$ and $\alpha_k<1$. Further, under conditions R2, R3, $\widetilde F_U$ is strictly increasing and continuously differentiable on $\{\alpha: (\alpha_{i_1}, \cdots,  \alpha_{i_p})\in\mathscr S_{\alpha_{i_1}(X),\cdots, \alpha_{i_p}(X)}; \text{other }\alpha_{i_j}=1\}$. Hence, we can extend it to the whole support $[0,1]^I$ as a proper distribution function such that it is strictly  increasing and continuously differentiable on $[0,1]^I$. The extended $\widetilde F_{U}(\cdot)$ will yield a positive and continuous conditional Radon--Nikodym density on $[0,1]^I$. 

By construction, $[\tilde \pi; \widetilde F_{U}]\in\mathcal M_2$. Fix $X=x$. The constructed structure $[\tilde \pi; \tilde F_{U}(\cdot)]$ will generate the given marginal distribution $\alpha_i(x)$ for all $i$. Moreover, for any tuple $\{ i_1,\cdots, i_p\}$ from $\{1,\cdots,I\}$,
\begin{multline*}
\widetilde \Pr(Y_{i_1}=1,\cdots, Y_{i_p}=1|X=x)=\widetilde F_{U_{i_1},\cdots, U_{i_p}}\left(\alpha_{i_1}(x), \cdots, \alpha_{i_p}(x)\right) \\
=\mathbb E \Big[\prod_{j=1}^p Y_{i_j}\big|\alpha_{i_1}(X)=\alpha_{i_1}(x),\cdots,\alpha_{i_p}(X)=\alpha_{i_p}(x)\Big]=\mathbb E \Big[\prod_{j=1}^p Y_{i_j}\big|X=x\Big].
\end{multline*}Because the tuple $\{ i_1,\cdots, i_p\}$ is arbitrary, then $[\tilde \pi, \widetilde F_U]$ generates the distribution $F_{Y|X}(\cdot|x)$.
\qed

\subsection{Proof of \Cref{ration_M_4}}
\label{proof_ration_M_4}
\proof
The ``only if part'' follows directly from assumption I and the single equilibrium condition. It suffices to show the ``if part''.

Fix a distribution $F_{Y|X}$ that satisfies the condition. Let $\tilde F_{U_i|X}=\tilde F_{U_i}$ be a uniform distribution on $[0,1]$ and $\tilde F_{U|X}=\prod_{i=1}^I\tilde F_{U_i}$. Moreover,  let $\tilde \pi_i(\cdot,x)=\alpha_i(x)$ for any $x\in\mathscr S_X$. By construction,  $[\tilde \pi;\tilde F_{U|X}]$ satisfies assumptions R, M, E, and I. Hence, $[\tilde \pi;\tilde F_{U|X}]\in\mathcal M_3$.

It suffices to show that the constructed structure $[\tilde \pi; \tilde F_{U|X}]$ can generate $F_{Y|X}$. Fix $x$. By construction, we have that $\widetilde \Pr(Y_i=1|X=x)=\alpha_i(x)$. Moreover,  for any tuple $ \{i_1,\cdots, i_p\}$ from $\{1,\cdots,I\}$, 
\begin{multline*}
\widetilde \Pr(Y_{i_1}=1,\cdots, Y_{i_p}=1|X=x)=\widetilde F_{U_{i_1},\cdots, U_{i_p}}\left(\alpha_{i_1}(x), \cdots, \alpha_{i_p}(x)\right) \\
=\prod_{j=1}^p \alpha_{i_j}(x)= \Pr(Y_{i_1}=1,\cdots, Y_{i_p}=1|X=x).
\end{multline*}Because the tuple $\{ i_1,\cdots, i_p\}$ is arbitrary, then $[\tilde \pi, \widetilde F_{U|X}]$ generates the distribution $F_{Y|X}(\cdot|x)$.
\qed

\section{Identification}
\subsection{Proof of \Cref{iden0}}
\label{app2}
\proof
Our proof is an extension of the copula argument in \cite{darsow1992copulas}. Fix $X=x$.  By law of iterated expectation, 
\begin{eqnarray*}
&&\Pr \left(Y_i=1; Y_{-i}=a_{-i}|\alpha(X)=\alpha \right)\\
&=&\mathbb E_{U_i} \left[\Pr \left(Y_i=1; Y_{-i}=a_{-i}|\alpha(X)=\alpha,U_i \right)\right]\\
&=&\int_{ Q_{U_i}(0)}^{Q_{U_i}(\alpha_i)} \Pr \left(Y_{-i}=a_{-i}|\alpha(X)=\alpha,U_i=u_i \right) dF_{U_i}(u_i)\\
&=&\int_{ 0}^{\alpha_i} \Pr \left[Y_{-i}=a_{-i}|\alpha(X)=\alpha,U_i=Q_{U_i}(v_i) \right] dv_i\\
&=&\int_{ 0}^{\alpha_i} \Pr \left[Y_{-i}=a_{-i}|\alpha_{-i}(X)=\alpha_{-i},U_i=Q_{U_i}(v_i) \right] dv_i
\end{eqnarray*}where the second equality comes from assumption E and the fact that $
\Pr[Y_i=1|\alpha(X)=\alpha, U_i\leq Q_{U_i}(\alpha_i)]=1$ and $
\Pr[Y_i=1|\alpha(X)=\alpha, U_i>Q_{U_i}(\alpha_i)]=0$,  the third equality  from a change--in--variable ($v_i=F_{U_i}(u_i)$) in the integration, and the last step is because conditioning on $\alpha_{-i}(X)$ is equivalent to conditioning on $u^*_{j}(X)$ for all $j\neq i$, therefore, $Y_{-i}$ is (conditionally) independent of $X$ (and $\alpha_i(X)$ as well) given $\alpha_{-i}(X)$ by assumption E.

Therefore, we have 
\[
\frac{\partial \Pr \left(Y_i=1; Y_{-i}=a_{-i}|\alpha(X)=\alpha \right)}{\partial \alpha_i}
=\Pr \left[Y_{-i}=a_{-i}|\alpha(X)=\alpha,U_i=Q_{U_i}(\alpha_i) \right].
\] Note that $Q_{U_i}(\alpha_i(x))=u^*_i(x)$. By assumption E, we then have
\begin{multline*}
\sigma^*_{-i}(a_{-i}|x,u^*_i(x))\equiv\Pr \left[Y_{-i}=a_{-i}|X=x,U_i=u^*_i(x) \right]\\
=\Pr \left[Y_{-i}=a_{-i}|\alpha(X)=\alpha(x),U_i=u^*_i(x) \right]
=\frac{\partial \Pr \left(Y_i=1; Y_{-i}=a_{-i}|\alpha(X)=\alpha \right)}{\partial \alpha_i}\Big|_{\alpha=\alpha(x)}.\qed
\end{multline*}

\subsection{Proof of \Cref{idco}}
\label{proof:idco}
\proof
By the proof in \Cref{lemma1} and  assumption ER, we have:  for all $x\in\mathscr S_X$ such that $\alpha_i(x)\in(0,1)$, 
\begin{equation}
\label{s_index}
\sum_{a_{-i}\in\mathcal{A}_{-i}} \pi_i(a_{-i},x_i)\sigma^*_{-i}\big(a_{-i}|x,u^*_i(x)\big)=Q_{U_i}(\alpha_i(x)).
\end{equation}
 It follows that
\begin{equation}
\label{s_index_expect}
\sum_{a_{-i}\in\mathcal{A}_{-i}} \pi_i(a_{-i},x_i)\mathbb E \left[\sigma^*_{-i}\big(a_{-i}|X,u^*_i(X))\big)|X_i=x_i,\alpha_i(X)=\alpha_i(x)\right]=Q_{U_i}(\alpha_i(x))
\end{equation}
The difference between (\ref{s_index}) and (\ref{s_index_expect}) yields
\begin{equation}
\label{s_index_2}
\sum_{a_{-i}\in\mathcal A_{-i}} \pi_i(a_{-i},x_i)\times\overline\sigma^*_{-i}\big(a_{-i},x\big)=0
\end{equation}
where $\overline\sigma^*_{-i}\big(a_{-i},x\big)\equiv \sigma^*_{-i}\big(a_{-i}|x,u^*_i(x)\big)-\mathbb E \left[\sigma^*_{-i}\big(a_{-i}|X,u^*_i(X)\big)|X_i=x_i,\alpha_i(X)=\alpha_i(x)\right]$.

When $x_i$ is fixed, we can identify $\pi_i(\cdot,x_i)$ as coefficients by varying $\overline\sigma^*_{-i}(a_{-i},x)$ through $x_{-i}$.  Suppose ${\mathcal R}_i(x_i)$  has rank $2^{I-1}-1$. Because $\sum_{a_{-i}\in\mathcal A_{-i}} \overline\sigma^*_{-i}\big(a_{-i},x\big)=0$,  $\pi_i(a_{-i},x_i)$ equals  the same constant for all $a_{-i}\in\mathcal A_{-i}$. By (\ref{s_index}), we have  $\pi_i(\cdot,x_i)=Q_{U_i}(\alpha_i(x))$. Therefore, $\mathscr S_{\alpha_i(X)|X_i=x_i}$ has to be a singleton $\{\alpha_i^\dag\}$.

Next, suppose  ${\mathcal R}_i(x_i)$  has rank $2^{I-1}-2$. Then we can pick a vector $\pi^0_i(\cdot,x_i)\in\mathbb R^{2^{I-1}}$ such that $\pi^0_i(a_{-i},x_i)\neq \pi^0_i(a_{-i}', x_i)$ for some $a_{-i}, a_{-i}'\in\mathcal A_{-i}$, and $\pi^0_i(\cdot,x_i)$ satisfy 
\[
\sum_{a_{-i}\in\mathcal A_{-i}}\pi^0_i(a_{-i},x_i)\times\overline\sigma^*_{-i}\big(a_{-i},x\big)=0.
\] Note that we also have $\sum_{a_{-i}\in\mathcal A_{-i}}1\times \overline\sigma^*_{-i}\big(a_{-i},x\big)=0$. 
By linear algebra, $\pi_i$ can be written as
\[
\pi_i(\cdot,x_i)= c_i(x_i) + p_i(x_i)\times \pi^0_i(\cdot,x_i) 
\] where $c_i,p_i:\mathscr S_{X_i}\rightarrow \mathbb R$. Hence, $\pi_i$ are identified up to location ($c_i$) and scale ($p_i$). 
\qed

\subsection{Proof of \Cref{idco_new}}
\label{proof:idco_new}
For the first half for the proposition, we show by contradiction. It is straightforward that $\mathscr S_{\alpha_i(X)|X_i=x_i}$ has to be a singleton if $\pi_i(\cdot,x_i)$ is constant on $\mathcal A_{-i}$.

We now show the identification of the sign of $\pi_i(a_i,x_i) -\pi_i(a_i^0,x_i)$. Let $x,x'\in\mathscr S_{X|X_i=x_i}$, $\alpha(x)=\alpha$, $\alpha(x')=\alpha'$, and w.l.o.g., $\alpha'_i<\alpha_i$. Then 
\[
\sum_{a_{-i}\in\mathcal{A}_{-i}} \pi_i(a_{-i},x_i)\sigma^*_{-i}\big(a_{-i}|x',Q_{U_i}(\alpha_i)\big)<Q_{U_i}(\alpha_i(x))=\sum_{a_{-i}\in\mathcal{A}_{-i}} \pi_i(a_{-i},x_i)\sigma^*_{-i}\big(a_{-i}|x,Q_{U_i}(\alpha_i)\big),
\] from which we have
\[
p_i(x_i)\times \sum_{a_{-i}\in\mathcal{A}_{-i}} \pi^0_i(a_{-i},x_i)\times\left[\sigma^*_{-i}\big(a_{-i}|x',Q_{U_i}(\alpha_i)\big)-\sigma^*_{-i}\big(a_{-i}|x,Q_{U_i}(\alpha_i)\big)\right]<0.
\]Thus we identify the sign of $p_i(x_i)$. It follows that the sign of $\pi_i(a_{-i},x_i)-\pi_i(a'_{-i},x_i)=p_i(x_i)\times\left[\pi^0_i(a_{-i},x_i)-\pi^0_i(a'_{-i},x_i)\right]$ is also identified. 
\qed

\subsection{Proof of \Cref{idpayoff}}
\label{proof_idpayoff}
\proof
By (\ref{eqcond}) and assumption N, clearly $\pi_i(a_{-i},\cdot)$  is identified  on   $\mathbb C^{0}_i$. Hence, it suffices to show that the identification of $\pi_i(a_{-i},\cdot)$    on   $\mathbb C^{t}_i$  implies its identification on  $\mathbb C^{t+1}_i$. By Definition 2, it suffices to consider $x_i\in\mathbb C^{t+1}_i/\mathbb C^t_i$. 

Suppose that Case (ii) occurs, i.e. ${\mathcal R}_i(x_i)$ has rank $2^{I-1}-2$  and there exists $x'_i\in  \mathbb C^{t}_i$ such that $\mathscr S_{\alpha_i(X)|X_i=x_i}\bigcap\mathscr S_{\alpha_i(X)|X_i=x'_i}\bigcap (0,1)$ contains at least two different elements $0<\alpha_i'<\alpha_i<1$. Let $x,x'\in\mathscr S_{X|X_i=x_i}$, $\alpha_i(x)=\alpha_i$ and $\alpha_i(x')=\alpha_i'$. Because $x'_i\in  \mathbb C^{t}_i$, then by assumption $\pi_i(\cdot,x'_i)$ are identified. Then both $Q_{U_i}(\alpha_i)$ and $Q_{U_i}(\alpha'_i)$ are identified by (\ref{eqcond}). Further, because ${\mathcal R}_i(x_i)$ has  rank $2^{I-1}-2$, then by \Cref{idco}, $\pi_i(\cdot,x_i)$ is identified up to location and scale, i.e. $\exists \ c_i(x_i), p_i(x_i)\in\mathbb R$ and a known vector $\pi^0_i(\cdot,x_i)\in\mathbb R^{2^{I-1}}$, such that $\pi_i(\cdot,x_i)=c_i(x)+p_i(x)\times \pi^0_i(\cdot,x_i)$. Moreover, because $\alpha_i,\alpha_i'\in\mathscr S_{\alpha_i(X)|X_i=x_i}\bigcap (0,1)$, then we have 
\begin{gather*}
\sum_{a_{-i}\in\mathcal A_{-i}} \pi_i(a_{-i},x_i)\times \sigma^*_{-i}(a_{-i}|x,u^*_i(x))=Q_{U_i}(\alpha_i),\\
\sum_{a_{-i}\in\mathcal A_{-i}} \pi_i(a_{-i},x_i)\times \sigma^*_{-i}(a_{-i}|x',u^*_i(x'))=Q_{U_i}(\alpha_i').
\end{gather*} It follows that
\begin{gather*}
c_i(x_i)+p_i(x_i)\times \sum_{a_{-i}\in\mathcal A_{-i}} \pi^0_i(a_{-i},x_i)\times \sigma^*_{-i}(a_{-i}|x,u^*_i(x))=Q_{U_i}(\alpha_i),\\
c_i(x_i)+p_i(x_i)\times \sum_{a_{-i}\in\mathcal A_{-i}} \pi^0_i(a_{-i},x_i)\times \sigma^*_{-i}(a_{-i}|x',u^*_i(x'))=Q_{U_i}(\alpha_i').
\end{gather*} Note that $Q_{U_i}(\alpha_i)$, $Q_{U_i}(\alpha'_i)$, $\pi^0_i(\cdot,x_i)$, $\sigma^*_{-i}(a_{-i}|x,u^*_i(x))$ and $\sigma^*_{-i}(a_{-i}|x',u^*_i(x'))$ are all known terms. Because $Q_{U_i}(\alpha'_i)<Q_{U_i}(\alpha_i)$, the determinant of the equation system cannot be zero. Then we can identify $c_i(x_i)$ and $p_i(x_i)$ from the above two equations. Therefore, $\pi_i(\cdot,x_i)$ are identified. 

Suppose that Case (iii) occurs, i.e. ${\mathcal R}_i(x_i)$ has  rank $2^{I-1}-1$  and there exists $x'_i\in  \mathbb C^{t}_i$ such that  $\mathscr S_{\alpha_i(X)|X_i=x_i}\subseteq\mathscr S_{\alpha_i(X)|X_i=x'_i}\bigcap (0,1)$. By \Cref{idco}, $\pi_i(\cdot,x_i)$ is identified by $Q_{U_i}(\alpha_i)$, which is known since  $\mathscr S_{\alpha_i(X)|X_i=x_i}\subseteq\mathscr S_{\alpha_i(X)|X_i=x'_i}\bigcap (0,1)$. 
\qed

\section{Extensions}

\subsection{Proof of \Cref{lem_set_1}}\label{proof_lem_set_1}
\proof First, we construct a structure $\widetilde S\in\mathcal M_2$  such that (1) $\widetilde S$ has the marginal quantile functions $\left(\widetilde Q_{U_1},\cdots,\widetilde Q_{U_I}\right)$; (2) $\widetilde C_U(\cdot)=C_U(\cdot)$ on $[0,1]^I$; (3) for any $x\in\mathscr S_X$, $i$, and $a_{-i}\in\mathcal A_{-i}$, let $\widetilde \pi_i(a_{-i}, x)=\widetilde Q_{U_i}\left(\mathbb E (Y_i|X=x)\right)$. By construction, it is straightforward that assumptions R, M and E are satisfied. 

Now it suffices to verify the observational equivalence between $\widetilde S$ and $S$. Fix $x\in\mathscr S_X$. Note that in the structure $\widetilde S$ there is no strategic effects, then the equilibrium is: $\textbf 1 \left\{u_i\leq \widetilde Q_{U_i}\left(\mathbb E (Y_i|X=x)\right)\right\}$ for $i=1,\cdots,I$. Here we only verify the observational equivalence for action profile $(1,\cdots,1)$ and the proofs for other action profiles follow similarly:
\begin{multline*}
\widetilde \Pr(Y_{1}=1;\cdots; Y_{I}=1|X=x)=\widetilde C_U\left(\mathbb E (Y|X=x)\right)\\
= C_U\left(\mathbb E (Y|X=x)\right)=\Pr(Y_{1}=1;\cdots; Y_{I}=1|X=x).\qed
\end{multline*}

\subsection{Proof of \Cref{set_id_1}}\label{proof_set_id_1}
\proof
It is straightforward that $\pi \in \Theta_I(\{Q_{U_i}\}_{i=1}^I, C_U)$. For  sharpness, it suffices to show that for any $\tilde \pi\in\Theta_I(\{\tilde Q_{U_i}\}_{i=1}^I, \tilde C_U)$, then  $\tilde S\equiv (\tilde \pi,\{\tilde Q_{U_i}\}_{i=1}^I,\tilde C_U)$, which belongs to $\mathcal M_2$ by the definition of $\Theta_I(\{\tilde Q_{U_i}\}_{i=1}^I, \tilde C_U)$, is observationally equivalent to the underlying structure $S\equiv (\pi,\{Q_{U_i}\}_{i=1}^I,C_U)$. 

Fix $X=x$. It suffices to verify that $\delta^*=\Big(\textbf 1 \left\{u_1\leq \tilde Q_{U_1}(\alpha_1(x))\right\}, \cdots, \textbf 1 \left\{u_I\leq \tilde Q_{U_I}(\alpha_I(x))\right\}\Big)$ is a BNE solution for the constructed structure. Because $\tilde C_U\in \mathscr C_I$ and by the proof for \Cref{iden0}, 
\[
\tilde\Pr_{\delta^*}\left\{Y_{-i}=a_{-i}|X=x,U_i= \tilde Q_{U_i}(\alpha_i(x))\right\}= \sigma^*_{-i}(a_{-i}|x,u^*_i(x)).
\] Then, by the conditions in the definition of $\Theta_I(\{\tilde Q_{U_i}\}_{i=1}^I, \tilde C_U)$, $\textbf 1 \left\{u_i\leq \tilde Q_{U_i}(\alpha_i(x))\right\}$ is the best response to $\delta^*_{-i}$. Thus $\delta^*$ is a BNE.
 \qed

\end{document}